\documentclass[english]{article}
\usepackage[T1]{fontenc}
\usepackage[latin9]{inputenc}
\usepackage{geometry}
\usepackage{babel}
\usepackage{array}
\usepackage{float}
\usepackage{mathrsfs}
\usepackage{multirow}
\usepackage{amsmath}
\usepackage{amssymb}
\usepackage{stmaryrd}
\usepackage{graphicx}
\usepackage{rotating}
\usepackage{setspace}
\usepackage[authoryear]{natbib}
\usepackage[unicode=true]
 {hyperref}

\makeatletter

\providecommand{\tabularnewline}{\\}
\floatstyle{ruled}
\newfloat{algorithm}{tbp}{loa}
\providecommand{\algorithmname}{Algorithm}
\floatname{algorithm}{\protect\algorithmname}

\usepackage{comment}
\usepackage{algpseudocode}
\usepackage{listings}
\usepackage{color}

\definecolor{dkgreen}{rgb}{0,0.6,0}

\makeatother

\begin{document}
\title{Connecting the Dots: Numerical Randomized Hamiltonian Monte Carlo
with State-Dependent Event Rates }
\author{Tore Selland Kleppe\thanks{Department of Mathematics and Physics, University of Stavanger, 4036
Stavanger, Norway. Email: tore.kleppe@uis.no}}
\maketitle
\begin{abstract}
Numerical Generalized Randomized Hamiltonian Monte Carlo is introduced,
as a robust, easy to use and computationally fast alternative to conventional
Markov chain Monte Carlo methods for continuous target distributions.
A wide class of piecewise deterministic Markov processes generalizing
Randomized HMC (Bou-Rabee and Sanz-Serna, 2017) by allowing for state-dependent
event rates is defined. Under very mild restrictions, such processes
will have the desired target distribution as an invariant distribution.
Secondly, the numerical implementation of such processes, based on
adaptive numerical integration of second order ordinary differential
equations (ODEs) is considered. The numerical implementation yields
an approximate, yet highly robust algorithm that, unlike conventional
Hamiltonian Monte Carlo, enables the exploitation of the complete
Hamiltonian trajectories (hence the title). The proposed algorithm
may yield large speedups and improvements in stability relative to
relevant benchmarks, while incurring numerical biases that are negligible
relative to the overall Monte Carlo errors. Granted access to a high-quality
ODE code, the proposed methodology is both easy to implement and use,
even for highly challenging and high-dimensional target distributions.
\end{abstract}

\section{Introduction}

By now, Markov chain Monte Carlo (MCMC) methods and their widespread
application in Bayesian statistics need no further introduction \citep[see e.g.][]{GelmanBDA3,robert_casella}.
In this paper, Generalized Randomized Hamiltonian Monte Carlo (GRHMC),
a wide class of continuous time Markov processes with a pre-selected
stationary distribution is constructed. Further, Numerical GRHMC (NGRHMC),
the practical implementation of GRHMC processes is suggested as a
robust and easy to use general purpose class of algorithms for solving
problems otherwise handled using MCMC methods for continuous state
spaces. 

The paper makes several contributions. GRHMC processes, a wide class
of piecewise deterministic Markov processes (PDMP) \citep[see e.g.][and references therein]{davis_PDP_book,fearnhead2018,1707.05296}
with target distribution-preserving Hamiltonian deterministic dynamics
are defined. GRHMC processes generalizes Randomized HMC \citep{bou-rabee2017}
by admitting state-dependent event-rates, while still retaining arbitrary
pre-specified stationary distributions. The highly flexible specification
of event rates leaves substantial room to construct processes that
are more optimized towards MCMC applications. A further benefit of
using conserving Hamiltonian deterministic dynamics is that GRHMC
processes are likely to scale well to high-dimensional problems (see
e.g., \citealp{2009.14239} where dimension-free convergence bounds
are obtained for the Anderson Thermostat, which generalizes RHMC).

Secondly, it is proposed to use adaptive numerical methods for second
order ordinary differential equations (ODEs) \citep[see e.g.][]{10.5555/153158}
to approximate a selected GRHMC process to arbitrary precision, leading
to NGRHMC. In common implementations of Hamiltonian Monte Carlo, errors
introduced by fixed time step (i.e., non-adaptive) symplectic/time-reversible
integrators are exactly corrected using accept/reject steps. Here,
biases relative to the (on target, but generally intractable) GRHMC
process stemming from the numerical integration of ODEs are not explicitly
corrected for, but rather kept under control by choosing the ODE integrator
precision sufficiently high. Numerical experiments indicate that even
for rather lax integrator precision, biases incurred by the numerical
integration of ODEs are imperceivable relative to the overall Monte
Carlo variation stemming from using MCMC-like methods. 

By allowing for such small biases, one may leverage high quality adaptive
ODE integration codes, making the proposed method both easy to implement
and requiring minimal expertise by the user. The system of ODEs may
be augmented beyond Hamilton's equations so that sampling of event
times under non-trivial event rates (without specifying model-specific
bounds on the event rates \citep{fearnhead2018}) and computing averages
over continuous time trajectories are done within ODE solver. This
practice of augmenting the ODE system ensures that the precision of
the overall algorithm (including the simulation of event times and
computation of moments) relative to the underlying GRHMC process is
controlled by a single error control mechanism and only a few easily
interpretable tuning parameters. Automatic tuning methodology of the
parameters in underlying GRHMC process, again leveraging aspects of
the adaptive ODE integration code, is also proposed.

Thirdly, it is demonstrated that certain moments of the target distribution
may be estimated extremely efficiently by exploiting the between events
Hamiltonian dynamics (and hence the name of the paper). Such improved
efficiency occur when the temporal averages of the position coordinate
of the Hamiltonian trajectories (without momentum refreshes) coincide
with the corresponding means under the target distributions. Such
effects occurs e.g., when estimating the mean under Gaussian target
distributions, but is by no means restricted to this situation. Exploitation
of any such effects is straight forward when the theoretical processes
are approximated using ODE solvers as proposed.

Finally, it is demonstrated that even rudimentary versions of NGRHMC
have competitive performance compared to commonly used (fixed time
step) symplectic/time-reversible methods for target distributions
where the latter methods work well. Furthermore, it is demonstrated
that adaptive nature of the integrators employed here resolves the
slow exploration associated with fixed step size HMC-MCMC chains for
target distributions exhibiting certain types of non-linearities.

This paper contains only initial steps towards understanding and exploiting
the full potential of GRHMC processes and their numerical implementation
and should be read as an invitation to further work. On the theoretical
side, understanding the ergodicity of GRHMC processes beyond RHMC,
bounding the biases stemming from numerical integration and understanding
scaling in high dimensions would be natural next steps. Better exploiting
the possibilities afforded by the highly flexible state-dependent
event rates constitutes another major avenue for further work. Throughout
the text, further suggestions for continued research in several other
regards are also pointed out.

\subsection{Relation to other work}

Continuous time Markov processes involving Hamiltonian dynamics subject
to random updates of velocities at random times are by no means new.
In the molecular simulation literature, the Anderson Thermostat (AT)
\citep{doi:10.1063/1.439486} involves Hamiltonian dynamics with updating
of the momentum of randomly chosen particles according to Bolzmann-Gibbs
distribution marginals. RHMC may be seen as a special case of the
AT with only one particle \citep{2009.14239}. Uncorrected numerical
implementations of the AT, involving fixed time step reversible integrators
may be found in several molecular simulation packages such as GROMACS
\citep{ABRAHAM201519}. 

The theoretical properties of the RHMC (with constant event rate and
exact Hamiltonian dynamics) and AT have been extensively studied:
\citet{bou-rabee2017} develop geometric ergodicity of RHMC under
mild assumptions. Further, \citet{10.1002/cpa.20198,10.1007/s10955-007-9391-0}
develop ergodicity for both continuous time AT and its time-discretization
under different assumptions, and \citet{2009.14239} consider the
convergence of AT in Wasserstein distance. \citet{2007.14927} study
RHMC under the hypocoercivity framework. As described by \citet{bou-rabee2017},
there is also an interesting and fundamental connection between RHMC
and second order Langevin dynamics \citep[see e.g.][ and references therein]{pmlr-v75-cheng18a}
in that the same stochastic Lyapunov function may be used to prove
geometric ergodicity of both types of processes. 

Recently, PDMPs \citep[see e.g.][and references therein]{davis_PDP_book,fearnhead2018,1707.05296}
have received substantial interest as time-irreversible alternatives
to conventional MCMC methods. Most proposed PDMP-based alternatives
to MCMC, such as the Bouncy Particle Sampler \citep{doi:10.1080/01621459.2017.1294075}
and the Zig-Zag process \citep{bierkens2019} rely on linear deterministic
dynamics. Another PDMP-based sampling algorithm; The Boomerang Sampler
(BS) \citep{2006.13777} uses the explicitly solvable Hamiltonian
deterministic dynamics associated with Gaussian approximation to the
target. The BS was found to outperform the mentioned linear dynamics
PDMPs. AT, RHMC along with GRHMC are also PDMPs based on Hamiltonian
deterministic dynamics, but unlike the BS, the involved deterministic
dynamics preserves exactly the target distribution, hence affording
GRHMC substantial flexibility with respect to the selection of event
rates. Interestingly, \citet{1808.04299} shows that the RHMC process
is a scaling limit of the Bouncy Particle Sampler \citep{doi:10.1080/01621459.2017.1294075}.
RHMC processes (i.e. with exact Hamiltonian dynamics and constant
event rates) are also mentioned in the context of PDMPs by \citet[footnote 3]{1707.05296},
but no details are provided on how to implement such an algorithm
are provided. 

Inter-event time sampling for PDMPs based on numerical integration
and root-finding (and thereby bypassing the need for global bounds
on the event rate as is also done in this paper) is considered by
\citet{2003.03636}. However, their approach is based on the linear
Zig-Zag dynamics and uses other numerical techniques to obtain integrated
event rates. HMC-MCMC based on exact Hamiltonian dynamics is considered
by \citet{doi:10.1080/10618600.2013.788448}, but their approach is
restricted to truncated Gaussian distributions where such dynamics
may be found on closed form. Theoretical work for HMC-MCMC assuming
exact target-preserving Hamiltonian dynamics may be found in e.g.,
\citet{1708.07114,DBLP:conf/approx/ChenV19}. \citet{10.1214/19-BA1171}
consider recycling the intermediate integrator-steps in HMC-MCMC in
a manner related to the temporal averages considered here, and also
find substantial improvements in simulation efficiency in numerical
experiments. 

Unadjusted (and therefore generally biased) numerical approximations
to intractable theoretical processes for simulation purposes have
received much attention, with the widely used stochastic gradient
Langevin dynamics \citep{10.5555/3104482.3104568} being such an example.
In particular, both first- and second order Langevin dynamics, along
with their multiple integration step counterpart, generalized HMC
\citep{HOROWITZ1991247}, have been implemented in unadjusted manners
\citep[see e.g.][for an overview]{Leimkuhler_molecular}. In addition,
several theoretical papers consider unadjusted HMC-MCMC algorithms,
see e.g., \citet{pmlr-v89-mangoubi19a}, \citet{2009.08735} and \citet{2105.00887}.
In this strand of literature, also so-called collocation methods (which
are closely related to certain Runge Kutta methods \citep{10.5555/153158})
are applied by \citet{1812.06243} for solving Hamilton's equations,
but their approach is again based on (discrete time-)HMC-MCMC and
does not appear to leverage modern numerical ODE techniques. Finally,
the present use of adaptive step size techniques has similarities
to \citet{SJOS:SJOS12204}, but the latter algorithm is based on Langevin
processes and involves a Metropolis-Hastings adjustment step. 

The reminder of this paper is laid out as follows: Section \ref{sec:Background}
provides background and fixes notation. GRHMC processes are defined
and discussed in Section \ref{sec:Continuous-time-HMC}. The practical
numerical implementation of such processes is discussed in Section
\ref{sec:Numerical-implementation}. Numerical experiments and illustrations,
along with benchmarks against Stan are given in Sections \ref{sec:Numerical-experiments}
and \ref{subsec:Dynamic-Inverted-Wishart}. Finally, Section \ref{sec:Discussion}
provides discussion. The complete set of source code underlying this
paper is available at \href{https://github.com/torekleppe/PDPHMCpaperCode}{https://github.com/torekleppe/PDPHMCpaperCode}.

\section{Background\label{sec:Background}}

This section provides some background and fixes notation for subsequent
use. Throughout this paper, a target distribution with density $\pi(\mathbf{q}),\;\mathbf{q}\in\Omega\subseteq\mathbb{R}^{d}$
with an associated density kernel $\tilde{\pi}(\mathbf{q})$ which
can be evaluated point-wise. The gradient/Jacobian operator of a function
with respect to some variable, say $\mathbf{x}$, is denoted by $\nabla_{\mathbf{x}}$.
Time-derivatives are denoted using the conventional dot-notation,
i.e. $\dot{f}(\tau)=\frac{d}{d\tau}f(\tau),$ $\ddot{f}(\tau)=\frac{d^{2}}{d\tau^{2}}f(\tau)$
for some function $f(\tau)$ evolving over time $\tau$. 

In the reminder of this section, Hamiltonian mechanics, HMC and PDMPs
are briefly reviewed in order to fix notation and provide the required
background. The reader is referred to \citet{goldstein2002classical,Leimkuhler:2004}
and \citet{1206.1901,bou-rabee_sanz-serna_2018} for more detailed
expositions of Hamiltonian mechanics and HMC. \citet{doi:10.1111/j.2517-6161.1984.tb01308.x,davis_PDP_book}
for consider PDMPs in general and \citet{fearnhead2018,1707.05296}
give details for Monte Carlo applications of PDMPs.

\subsection{Elements of Hamiltonian mechanics}

Hamiltonian Monte Carlo methods rely on specifying a physical system
and use the dynamics of this system to propose transitions. The state
$\mathbf{z}=[\mathbf{q}^{T},\mathbf{p}^{T}]^{T}\in\mathbb{R}^{2d}$
of the system is characterized by a position coordinate $\mathbf{q}\in\mathbb{R}^{d}$
and a momentum coordinate $\mathbf{p}\in\mathbb{R}^{d}$. The system
itself is conventionally specified in terms of the Hamiltonian $\mathcal{H}(\mathbf{z})=\mathcal{H}(\mathbf{q},\mathbf{p})$
which gives the total energy of the system for a given state $\mathbf{z}$.
Throughout this work, physical systems with Hamiltonian given as 
\begin{equation}
\mathcal{H}(\mathbf{q},\mathbf{p})=-\log\tilde{\pi}(\mathbf{q})+\frac{1}{2}\mathbf{p}^{T}\mathbf{M}^{-1}\mathbf{p},\label{eq:Hamiltonian}
\end{equation}
are considered. Here $\mathbf{M}\in\mathbb{R}^{d\times d}$ is a symmetric,
positive definite (SPD) mass matrix which is otherwise specified freely.
The time-evolution of the system is given by Hamilton's equations
$\dot{\mathbf{q}}(\tau)=\nabla_{\mathbf{p}}\mathcal{H}(\mathbf{q}(\tau),\mathbf{p}(\tau)),$
$\dot{\mathbf{p}}(\tau)=-\nabla_{\mathbf{q}}\mathcal{H}(\mathbf{q}(\tau),\mathbf{p}(\tau))$,
which for Hamiltonian (\ref{eq:Hamiltonian}) reduces to:
\begin{equation}
\dot{\mathbf{z}}(\tau)=\left[\begin{array}{c}
\dot{\mathbf{q}}(\tau)\\
\dot{\mathbf{p}}(\tau)
\end{array}\right]=\left[\begin{array}{c}
\mathbf{M}^{-1}\mathbf{p}(\tau)\\
\nabla_{\mathbf{q}}\log\tilde{\pi}(\mathbf{q}(\tau))
\end{array}\right]\label{eq:Ham_eq}
\end{equation}
The flow associated with (\ref{eq:Ham_eq}) is denoted by $\varphi_{\tau}(\cdot)$,
and is defined so that $\mathbf{z}(\tau+s)=\varphi_{s}(\mathbf{z}(\tau))$
solves (\ref{eq:Ham_eq}) for any scalar time increment $s$, initial
time $\tau$ and initial configuration $\mathbf{z}(\tau)$. The flow
can be shown to be 
\begin{itemize}
\item Energy preserving, i.e. $\frac{\partial}{\partial\tau}\mathcal{H}(\varphi_{\tau}(\mathbf{z}))=0$
for all admissible $\mathbf{z}$, 
\item Volume preserving, i.e. $|\nabla_{\mathbf{z}}\varphi_{\tau}(\mathbf{z})|=1$
for each fixed $\tau$ and all admissible $\mathbf{z}$, 
\item Time reversible, which in the present context is most conveniently
formulated via that $\mathbf{T}_{\tau}=\mathbf{R}\circ\varphi_{\tau}$
is an involution so that $\mathbf{T}_{\tau}\circ\mathbf{T}_{\tau}$
is the identity operator. The momentum flip operator $\mathbf{R}=\text{diag}(\mathbf{I}_{d},-\mathbf{I}_{d})$
effectively reverses time.
\end{itemize}

\subsection{Hamiltonian Monte Carlo\label{subsec:Hamiltonian-Monte-Carlo}}

\begin{figure}
\centering{}\includegraphics[scale=0.55]{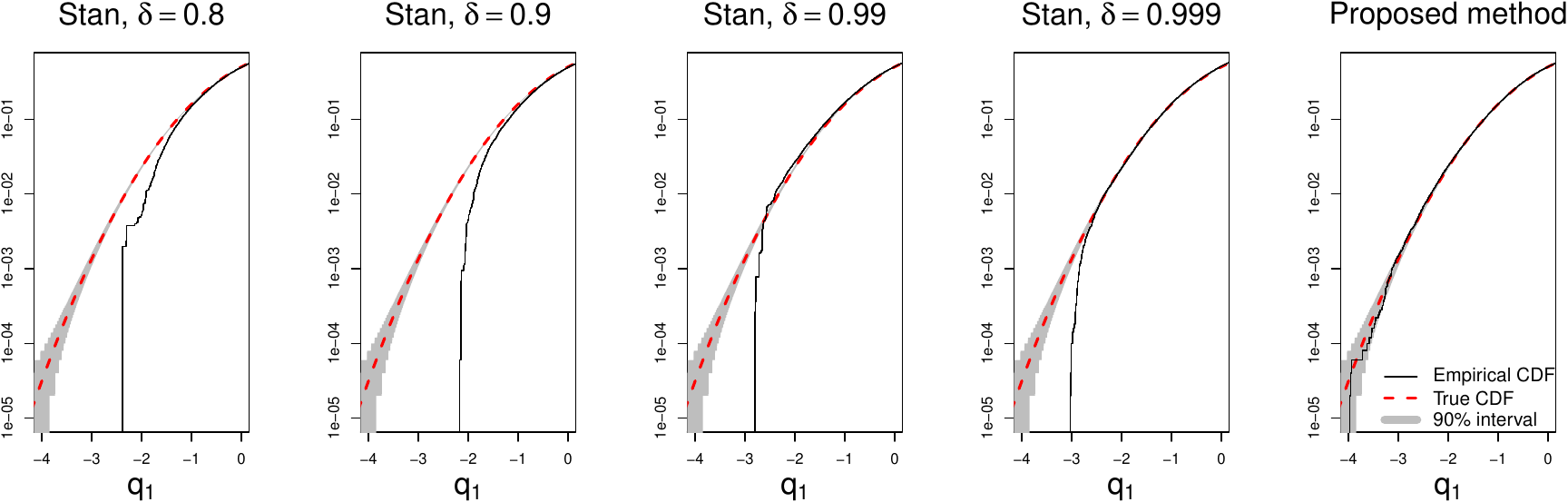}\caption{\label{fig:Initial-experiment-funnel}Empirical cumulative distribution
functions (CDFs) associated with MCMC output for the standard Gaussian
$\mathbf{q}_{1}$-marginal under the ``funnel''-model $\mathbf{q}_{1}\sim N(0,1)$,
$\mathbf{q}_{2}|\mathbf{q}_{1}\sim N(0,\exp(3\mathbf{q}_{1}))$. For
visual clarity, only the left half of the distributions are presented.
Further details on this experiment can be found in Section \ref{subsec:Funnel-distribution}.
Each case is based on 5000 samples from each of 10 independent replica.
The black solid lines are the empirical CDFs, whereas the red dashed
lines are the true CDFs. The shaded gray region would cover 90\% of
empirical CDFs based on 50000 iid $N(0,1)$ samples point-wise. The
four left-most panels are based on Stan output with different values
of the accept rate target $\delta$. In practice, higher values of
$\delta$ corresponds to smaller integrator step sizes and more integrator
steps per produced sample. The right-most panel shows output for the
proposed methodology using constant event rates. For both Stan and
the proposed methodology, an identity mass matrix was employed.}
\end{figure}
In the context of statistical computing, Hamiltonian dynamics has
attracted much attention the last decade. This interest is rooted
in that the flow $\varphi_{\tau}$ of (\ref{eq:Ham_eq}) (and associated
involution $\mathbf{T}_{\tau}$) exactly preserves the Boltzmann-Gibbs
(BG) distribution 
\begin{equation}
\rho(\mathbf{z})=\rho(\mathbf{q},\mathbf{p})=\pi(\mathbf{q})\mathcal{N}(\mathbf{p}|\mathbf{0}_{d},\mathbf{M})\propto\exp(-\mathcal{H}(\mathbf{q},\mathbf{p})),\label{eq:Boltzmann}
\end{equation}
associated with $\mathcal{H}$. I.e., for each fixed time increment
$\tau$,\emph{ $\varphi_{\tau}(\mathbf{z})$ $\sim\rho$ whenever
$\mathbf{z}\sim\text{\ensuremath{\rho}}$.} It is seen that the target
distribution is the $\mathbf{q}$-marginal of the BG distribution.
Thus, a hypothetical MCMC algorithm targeting (\ref{eq:Boltzmann}),
and producing samples $\mathbf{z}_{(i)}=(\mathbf{q}_{(i)}^{T},\mathbf{p}_{(i)}^{T})^{T}$,
would involve the BG distribution-preserving steps:
\begin{itemize}
\item Sample $\mathbf{p}_{*}\sim N(\phi\mathbf{p}_{(i-1)},\sqrt{1-\phi^{2}}\mathbf{M})$
for some $\phi\in(-1,1)$ and set $\mathbf{z}_{*}=(\mathbf{q}_{(i-1)}^{T},\mathbf{p}_{*}^{T})^{T}$.
\item For some suitable time increment $s$, $\mathbf{z}_{(i)}=\varphi_{s}(\mathbf{z}_{*})$.
\end{itemize}
Subsequently, the momentum samples, $\mathbf{p}_{(i)}$, may be discarded
to obtain samples targeting $\pi(\mathbf{q})$ only. Randomized durations
$s$ may be introduced in order to avoid periodicities or near-periodicities
in the underlying dynamics, and may result in faster convergence \citep{MACKENZE1989369,MZA:8194876,bou-rabee2017,bou-rabee_sanz-serna_2018}.
Alternatively, more sophisticated algorithms can be used to avoid
u-turns \citep{JMLR:v15:hoffman14a}. 

For all but the most analytically tractable target distributions,
the flow associated with Hamilton's equations is not available in
closed form, and hence it must be integrated numerically for the practical
implementation of the above MCMC sampler. Provided a time-reversible
integrator is employed for this task, the numerical error incurred
in the second step of the above MCMC algorithm can be exactly corrected
using an accept/reject step \citep{doi:10.1063/1.4874000}, but the
accept probability may be computationally demanding. If the employed
integrator is also volume preserving (i.e. symplectic) \citep[see e.g.][]{Leimkuhler:2004},
the accept-probability depends only on the values of the Hamiltonian
before and after the integration. This simplification has led to the
widespread application of the symplectic leap-frog (or Størmer-Verlet)
integrator in HMC implementations such as e.g., Stan \citep{stan-manual}.

Requiring the integrator to be time reversible and symplectic imposes
rather strict restrictions on the integration process. In particular
the application of adaptive (time-)step sizes, which is an integral
part of any modern general purpose numerical ODE code, is at best
difficult to implement \citep[see][Chapter 9 for discussion of this problem]{Leimkuhler:2004}
while maintaining time-reversible and symplectic properties. 

Figure \ref{fig:Initial-experiment-funnel} illustrates the effect
of using fixed step sizes for the funnel-type distribution $q_{1}\sim N(0,1)$,
$q_{2}|q_{1}\sim N(0,\exp(3q_{1}))$ (further details on this experiment
can be found in Section \ref{subsec:Funnel-distribution}). In the
four left panels, empirical cumulative distribution functions (CDFs)
calculated from MCMC output using the fixed step size integrator in
Stan are depicted for various accept rate targets $\delta$ which
has a default value of 0.8. In practice, higher values of $\delta$
correspond to higher fidelity integration with smaller integrator
step sizes and more integrator steps per produced sample. It is seen
that even with very small step sizes, Stan fails to properly represent
the left-hand tail of $q_{1}$ (which imply a very small scale in
the $q_{2}$). In the $\delta=0.999$ case, the smallest produced
sample out 50000 is $\approx-3.026$. In an iid $N(0,1)$ sample of
this size, one would expect around 62 samples smaller than this value. 

Also included in Figure \ref{fig:Initial-experiment-funnel} are results
from a variant of the proposed methodology (see Section \ref{subsec:Funnel-distribution}
for details), which is based on adaptive numerical integrators. The
method shows no such pathologies, and in particular the number of
samples below the smallest $\delta=0.999$ Stan sample was 70. 

\subsection{Piecewise deterministic Markov processes\label{subsec:Piecewise-deterministic-processe}}

Recently, continuous time piecewise deterministic Markov processes
(PDMP) \citep[see e.g.][]{doi:10.1111/j.2517-6161.1984.tb01308.x,davis_PDP_book}
have been considered as alternatives to discrete time Markov chains
produced by conventional MCMC methods. PDMPs may be employed for simulating
dependent samples, or more generally continuous time trajectories
with a given marginal probability distribution \citep[see][and references therein]{fearnhead2018}.
As the name indicates, PDMPs follow a deterministic trajectory between
events occurring at stochastic times. At events, the state is updated
in a stochastic manner. 

Following \citet{fearnhead2018}, a PDMP, say $\mathcal{Z}(t)\in\mathbb{R}^{D}$,
$t\in[0,\infty)$, is specified in terms of three components $(\Phi,\lambda,Q)$:
\begin{itemize}
\item Deterministic dynamics on \emph{time intervals where events do not
occur}, specified in terms of a set of ODEs: $\dot{\mathcal{Z}}(t)=\Phi(\mathcal{Z}(t))$.
\item A non-negative event rate $\lambda(\mathcal{Z}(t))$, depending only
on the current state of the process, so that the probability of an
event between times $t$ and $t+r$, $r\geq0$ is $\lambda(\mathcal{Z}(t))r+o(r)$
for small $r$.
\item Finally, a ``transition distribution at events'' $Q(\cdot|\mathcal{Z}(t-))$.
Suppose an event occurs at time $t$, and $\mathcal{Z}(t-)$ is the
state immediately before time $t$. Then the $\mathcal{Z}(t)$ will
be drawn randomly with density $Q(\cdot|\mathcal{Z}(t-))$.
\end{itemize}
Let $\Xi_{s}$ be the flow associated with $\Phi$. In order to simulate
from a PDMP, suppose first that $\mathcal{Z}(0)$ has been set to
some value, and that $t$ is initially set to zero. Then the following
three steps are repeated until $t>T$ where $T$ is the desired length
of the PDMP trajectory:
\begin{itemize}
\item Simulate a new $u\sim$ Exp$(1)$ and subsequently compute the time-increment
until next event $v$, which obtains as the solution in $v$ to 
\begin{equation}
\Lambda(v;\mathcal{Z}(t))=u,\;\text{where }\Lambda(v;\mathbf{z})=\int_{0}^{v}\lambda(\Xi_{s}(\mathcal{Z}(t)))ds.\label{eq:integrated_event_intensity}
\end{equation}
\item Set $\mathcal{Z}(t+s)=\Xi_{s}(\mathcal{Z}(t))$ for all $s\in[0,v)$,
and $\mathcal{Z}^{\text{*}}=\Xi_{v}(\mathcal{Z}(t))$. 
\item Set $t\leftarrow t+v$ and simulate $\mathcal{Z}(t)\sim Q(\cdot|\mathcal{Z}^{*})$. 
\end{itemize}
An invariant distribution of the process $\mathcal{Z}(t)$, say $p(\mathbf{z})$,
will satisfy the time-invariant Fokker-Planck/Kolmogorov forward equation
\citep{fearnhead2018}
\begin{equation}
\sum_{i=1}^{D}\frac{\partial}{\partial z_{i}}\left[\Phi_{i}(\mathbf{z})p(\mathbf{z})\right]=\int p(\mathbf{z}^{\prime})\lambda(\mathbf{z}^{\prime})Q(\mathbf{z}|\mathbf{z}^{\prime})d\mathbf{z}^{\prime}-p(\mathbf{z})\lambda(\mathbf{z}),\label{eq:fokker_planck}
\end{equation}
for all admissible states $\mathbf{z}$. For continuous time Monte
Carlo applications, one therefore seeks combinations of $(\Phi,\lambda,Q)$
so that the desired target distribution is an invariant distribution
$p(\mathbf{z})$. 

Provided such a combination has been found, discrete time Markovian
samples 
\begin{equation}
\mathbf{z}_{(i)}=\mathcal{Z}(\Delta i),\;\text{for some sample spacing \ensuremath{\Delta>0},}\label{eq:discrete_samples}
\end{equation}
may be used in the same manner as regular MCMC samples for characterizing
the invariant distribution. In addition, by letting the sample spacing
$\Delta\rightarrow0$, moments under the invariant distribution may
also be obtained by utilizing the complete trajectory of the PDMP,
i.e.
\begin{equation}
\frac{1}{T}\int_{0}^{T}g(\mathcal{Z}(t))dt\underset{T\rightarrow\infty}{\longrightarrow}\int g(\mathbf{z})p(\mathbf{z})d\mathbf{z}\;\text{almost surely},\label{eq:integrated-trajectory}
\end{equation}
for some function $g$. 

In most current implementations of PDMPs, $\lambda$ depends on $\nabla_{\mathbf{q}}\log\tilde{\pi}(\mathbf{q})$
and hence $\Lambda(v;\mathbf{z})$ cannot be evaluated analytically,
complicating the simulation of between event times, $v$, according
to (\ref{eq:integrated_event_intensity}). The between event times
are most commonly resolved using thinning \citep[see e.g.][Section 2.1]{fearnhead2018},
which in turn necessitates selecting an upper bound on $\lambda$
specific to the target distribution in question. The tightness of
the bound substantially influences the computational cost of the resulting
method. Similar to the present work, \citealt{2003.03636} bypasses
the need for such bounds by approximating $\Lambda(v;\mathbf{z})$
using numerical integration and solve (\ref{eq:integrated_event_intensity})
using numerical root finding, thereby obtaining a PDMP that is slightly
biased relative to the target distribution. 

\section{Generalized randomized HMC processes\label{sec:Continuous-time-HMC}}

In this section, theoretical PDMPs with (appropriately chosen) Hamiltonian
dynamics (\ref{eq:Ham_eq}) between events are considered. These processes
will be referred to a generalized randomized HMC processes (GRHMC),
and it shown that for a large class of combinations of $(\lambda,Q)$,
GRHMC processes will have the BG distribution (\ref{eq:Boltzmann})
as a stationary distribution. In practice, the Hamiltonian flow is
implemented using high precision adaptive numerical methods (to be
discussed in Section \ref{subsec:Numerical-method-for} and referred
to as Numerical GRHMC), to obtain a robust and accurate, but nevertheless
\emph{approximate} versions of these PDMPs.

\subsection{GRHMC as PDMPs\label{subsec:GRHMC-as-PDMPs}}

The GRHMC process targeting $\rho(\mathbf{z})$ is constructed within
the PDMP framework of Section \ref{subsec:Piecewise-deterministic-processe}
as follows; set $D=2d$, $\mathbf{z}=[\mathbf{q}^{T},\mathbf{p}^{T}]^{T}$
and,
\begin{itemize}
\item The deterministic dynamics are Hamiltonian, namely 
\begin{equation}
\Phi(\mathbf{z})=\left[\begin{array}{c}
\mathbf{M}^{-1}\mathbf{p}\\
\nabla_{\mathbf{q}}\log\tilde{\pi}(\mathbf{q})
\end{array}\right],\text{ and hence }\Xi_{\tau}=\varphi_{\tau}.\label{eq:PDMP_ham_dyn}
\end{equation}
\item A general \emph{state-dependent} event rate $\lambda(\mathbf{z})=\lambda(\mathbf{q},\mathbf{p})>0$
subject only to the restriction that $C(\mathbf{q})=\int\lambda(\mathbf{q},\mathbf{p})\mathcal{N}(\mathbf{p}|\mathbf{0}_{d},\mathbf{M})d\mathbf{p}<\infty$
(for all admissible $\mathbf{q}$) is assumed. 
\item The ``transition distribution at events'' is given in terms of the
density
\begin{equation}
Q(\mathbf{z}|\mathbf{z}^{\prime})=\delta(\mathbf{q}-\mathbf{q}^{\prime})K_{\mathbf{q}^{\prime}}(\mathbf{p}|\mathbf{p}^{\prime}),\label{eq:general_Q_paper-1}
\end{equation}
where $K_{\mathbf{q}}(\mathbf{p}|\mathbf{p}^{\prime})$ is a Markov
kernel density which leaves $v_{\mathbf{q}}(\mathbf{p})=\lambda(\mathbf{q},\mathbf{p})\mathcal{N}(\mathbf{p}|\mathbf{0}_{d},\mathbf{M})\left[C(\mathbf{q})\right]^{-1}$
invariant for all fixed $\mathbf{q}$, where and $\delta(\cdot)$
is the Dirac delta function centered in $\mathbf{0}$.
\end{itemize}
From now on, $\mathcal{Q}(t)$ and $\mathcal{P}(t)$ are used for
position- and momentum sub-vectors of $\mathcal{Z}(t)$ respectively,
i.e. $\mathcal{Z}(t)=[\mathcal{Q}(t)^{T},\mathcal{P}(t)^{T}]^{T},\;t\in[0,T]$.

\subsection{Stationary distribution}

\textbf{Proposition 1:} \emph{The above introduced GRHMC processes
admit $\rho(\mathbf{z})$ as a stationary distribution.}

Proposition 1 is proved by showing that both sides of the steady state
Fokker-Planck equation (\ref{eq:fokker_planck}) with $p(\mathbf{z})=\rho(\mathbf{z})$
are zero for a GRHMC process. As shown in Appendix \ref{subsec:Poisson_bracket},
for BG-preserving Hamiltonian dynamics (\ref{eq:PDMP_ham_dyn}) between
events, the left-hand side of the Fokker-Planck equation (\ref{eq:fokker_planck})
vanishes,\emph{ }i.e.
\begin{equation}
\sum_{i=1}^{D}\frac{\partial}{\partial z_{i}}\left[\Phi_{i}(\mathbf{z})\rho(\mathbf{z})\right]=0.\label{eq:fp_left_hand_side}
\end{equation}
Further, due to the $v_{\mathbf{q}}$-preserving nature of $K_{\mathbf{q}}(\mathbf{p}|\mathbf{p}^{\prime})$
above, the right hand side of the Fokker-Planck equation (\ref{eq:fokker_planck})
reduces to (see Appendix \ref{subsec:General-Markov-kernel} for more
detailed calculations)
\begin{align*}
 & \int\rho(\mathbf{z}^{\prime})\lambda(\mathbf{q}^{\prime},\mathbf{p}^{\prime})\delta(\mathbf{q}-\mathbf{q}^{\prime})K_{\mathbf{q^{\prime}}}(\mathbf{p}|\mathbf{p}^{\prime})d\mathbf{z}^{\prime}-\rho(\mathbf{z})\lambda(\mathbf{z}),\\
= & \pi(\mathbf{q})C(\mathbf{q})\int K_{\mathbf{q}}(\mathbf{p}|\mathbf{p}^{\prime})v_{\mathbf{q}}(\mathbf{p}^{\prime})d\mathbf{p}^{\prime}-\rho(\mathbf{z})\lambda(\mathbf{z})=0,
\end{align*}
and hence Proposition 1 follows.

Notice that allowing the event rate to depend on the momentum $\mathbf{p}$
requires that the momentum refresh distribution must be modified relative
to simply preserving the BG distribution $\mathbf{p}$-marginal as
in regular HMC and RHMC. Similar choices of $Q$ are discussed by
\citet[Section 3.2.1]{fearnhead2018} and \citet[Section 2.3.3]{1707.05296}.
Further notice that the above results are easily modified to accommodate
a general non-Gaussian $\mathbf{p}$-marginal \citep[see e.g.][]{10.1093/biomet/asz013}
of the (separable) BG distribution (see Appendix \ref{subsec:Poisson_bracket},\ref{subsec:General-Markov-kernel})
and a Riemann manifold variant \citep{girolami_calderhead_11} (see
Appendix \ref{sec:Riemann-manifold-variants}).

\subsection{More on event specifications}

Two special cases of the general event specification characterized
by $\lambda=\lambda(\mathbf{q},\mathbf{p})$ and (\ref{eq:general_Q_paper-1})
may be mentioned: for event rates not depending on $\mathbf{p}$,
say $\lambda(\mathbf{z})=\omega(\mathbf{q})>0$, implies that $v_{\mathbf{q}}(\mathbf{p})=\mathcal{N}(\mathbf{p}|\mathbf{0}_{d},\mathbf{M})\;\forall\;\mathbf{q}$
and momentums may be updated as in generalized HMC \citep{HOROWITZ1991247},
namely
\begin{equation}
K_{\mathbf{q}}(\mathbf{p}|\mathbf{p}^{\prime})=\mathcal{N}(\mathbf{p}|\phi\mathbf{p}^{\prime},\sqrt{1-\phi^{2}}\mathbf{M}),\label{eq:partial_refresh_lambda}
\end{equation}
for some fixed Horowitz parameter $\phi\in(-1,1)$. Secondly, assuming
further structure on $\lambda=\lambda(\mathbf{q},\mathbf{p})$ may
lead to tractable sampling directly from $v_{\mathbf{q}}(\mathbf{p})$.
Examples include: 
\begin{itemize}
\item $\lambda$ allows the representation $\lambda(\mathbf{z})=b(\mathbf{q},\mathbf{p}^{T}\mathbf{M}^{-1}\mathbf{p})$
for suitably chosen function $b:\mathbb{R}^{d}\times\mathbb{R}^{+}\mapsto\mathbb{R}^{+}$.
Then, $v_{\mathbf{q}}(\mathbf{p})$ is an elliptically contoured distribution
\citep[see e.g.][]{CAMBANIS1981368} which typically allows efficient
independent sampling.
\item $\log(\text{\ensuremath{\lambda}}(\mathbf{z}))$ is a quadratic function
in $\mathbf{p}$ for each $\mathbf{q}$. Then $v_{\mathbf{q}}(\mathbf{p})$
is Gaussian, which admit straight forward independent or autocorrelated
momentum refreshes.
\end{itemize}
These cases, and the rather rudimentary specific choices committed
below, are by no means exhausting the possibilities, and further research
taking (\ref{eq:general_Q_paper-1}) as vantage point is currently
under way. Avenues actively explored include Metropolized versions
of $K_{\mathbf{q}}(\mathbf{p}|\mathbf{p}^{\prime})$ similar to (\ref{eq:partial_refresh_lambda})
but for general $\mathbf{p}$-dependent event rates. Further work
is done to obtain processes that have intervals between events well
adapted to the target distribution similarly to e.g., the NUTS algorithm.
Finally, it should also be mentioned that $K_{\mathbf{q}}(\mathbf{p}|\mathbf{p}^{\prime})$
may in principle be selected first for some desirable purpose (e.g.,
momentum-refreshes that are over-dispersed relative to $\mathbf{M}$
to allow for jumps between modes), with the event rate subsequently
chosen to so that $K_{\mathbf{q}}$ is invariant with respect to $v_{\mathbf{q}}$. 

Further notice that there is a fundamental difference between changing
the BG-$\mathbf{p}$-marginal \citep[see e.g.][]{10.1093/biomet/asz013}
and selecting a $\mathbf{p}$-dependent event rate so that the moment-refreshes
must preserve a $v_{\mathbf{q}}(\mathbf{p})$ different from $\mathcal{N}(\mathbf{p}|\mathbf{0}_{d},\mathbf{M})$.
The former case changes also the deterministic dynamics, whereas the
latter does not. Hence, there is \emph{an additional degree of freedom}
in the present setup in that one may fix the Hamiltonian (and hence
dynamics) first, and then modify the momentum refreshes afterwards
by suitable choices of the event rate. 

\subsection{Specific event specifications}

\begin{table}
\centering{}%
\begin{tabular}{llll}
\hline 
Event  & \multirow{2}{*}{$\lambda$} & \multirow{2}{*}{$K_{\mathbf{q}}(\mathbf{p}|\mathbf{p}^{\prime})$} & \multirow{2}{*}{Interpretation}\tabularnewline
specification &  &  & \tabularnewline
\hline 
1 & $\frac{1}{\beta}$ & $\mathcal{N}(\mathbf{p}|\phi\mathbf{p}^{\prime},\sqrt{1-\phi^{2}}\mathbf{M}),\;\phi\in(-1,1)$ & Time between events is $Exp(\beta$), \tabularnewline
 &  &  & autocorrelated momentum refreshes\tabularnewline
 &  &  & \tabularnewline
2 & $\frac{1}{\beta}\sqrt{\mathbf{p}^{T}\mathbf{M}^{-1}\mathbf{p}}$ & $\propto\sqrt{\mathbf{p}^{T}\mathbf{M}^{-1}\mathbf{p}}\exp\left(-\frac{1}{2}\mathbf{p}^{T}\mathbf{M}^{-1}\mathbf{p}\right)$ & Arc-length of between-events\tabularnewline
 &  & $\sim\sqrt{\frac{r}{\mathbf{y}^{T}\mathbf{y}}}\sqrt{\mathbf{M}}\mathbf{y},$ & (standardized) position trajectory is $Exp(\beta)$,\tabularnewline
 &  & where $\mathbf{y}\sim N(\mathbf{0}_{d},\mathbf{I}_{d}),\;r\sim\chi^{2}(d+1)$ & independent momentum refreshes\tabularnewline
\hline 
\end{tabular}\caption{\label{tab:The-event-specifications}The event specifications applied
in the reminder of this text. In all cases $\beta$ is a tuning parameter,
where larger $\beta$s on average correspond to less frequent events/longer
inter-event trajectories. For specification 2, arc-lengths of position-trajectories
are calculated in the Mahalanobis distance $d(\mathbf{q},\mathbf{q}^{\prime})=\sqrt{(\mathbf{q}-\mathbf{q}^{\prime})^{T}\mathbf{M}(\mathbf{q}-\mathbf{q}^{\prime})}$
as $\mathbf{M}^{-1}$ is assumed to be some approximation/reflect
the scales of the covariance matrix of $\pi(\mathbf{q})$.}
\end{table}
Table \ref{tab:The-event-specifications} provides the specific event
specifications used in the remainder of this text. The former is RHMC
with Horowitz type momentum refreshes (\ref{eq:partial_refresh_lambda}),
whereas specification 2 involves independent updates according to
an elliptically contoured momentum refresh distribution $v_{\mathbf{q}}$.

Interestingly, the large $\lambda$ limit of the $\mathbf{q}$-component
of the PDMP, i.e. $\mathcal{Q}(t)$, for specification 1, is a Brownian
motion-driven preconditioned Langevin process \citep[see e.g.][]{RSSB:RSSB123}
(see Appendix \ref{sec:Langevin-limit})
\begin{equation}
d\mathcal{Q}(t)=\frac{1}{2}\mathbf{M}^{-1}\nabla_{\mathbf{q}}\log\tilde{\pi}(\mathcal{Q}(t))+\mathbf{M}^{-\frac{1}{2}}d\mathbf{W}(t).\label{eq:langevin_limit}
\end{equation}
Here $\mathbf{W}(t)$ is a standard Brownian motion and $\mathbf{M}^{-\frac{1}{2}}$
is any matrix square-root of $\mathbf{M}^{-1}$.

Specification 2 is a first attempt at providing event rates where
the length of the between-event trajectories is chosen dynamically.
Specifically, the event specification is chosen so that $\beta\int_{0}^{v}\lambda(\varphi_{s}(\mathbf{z}))ds=\beta\Lambda(v;\mathbf{z})$
is exactly the arc-length (in the Mahalanobis distance $d(\mathbf{q},\mathbf{q}^{\prime})=\sqrt{(\mathbf{q}-\mathbf{q}^{\prime})^{T}\mathbf{M}(\mathbf{q}-\mathbf{q}^{\prime})}$
for standardization, see Appendix Section \ref{subsec:Arc-length-and-event}
for details) of the position coordinate when $\mathbf{z}$ was the
state of the process immediately after the last event. For this specification,
$v_{\mathbf{q}}$ allows straight forward sampling as it is an elliptically
contoured distribution (see Table \ref{tab:The-event-specifications}). 

Note that $\beta=O(d^{1/2})$ is needed to ensure $E(\lambda)=O(1)$,
but also result in that $Var(\lambda)=O(d^{-1})$ under the BG-distribution.
Further, $v_{\mathbf{q}}$ converges to $N(\mathbf{0}_{d},\mathbf{M})$
for large $d$ (as $Var(r/\mathbf{y}^{T}\mathbf{y})$ in Table \ref{tab:The-event-specifications}
is $O(d^{-1})$). Hence, for large $d$ one would expect a similar
behavior of specifications 1 and 2.

\subsection{Illustrative examples\label{subsec:Illustrative-example}}

\begin{figure}
\centering{}\includegraphics[scale=0.55]{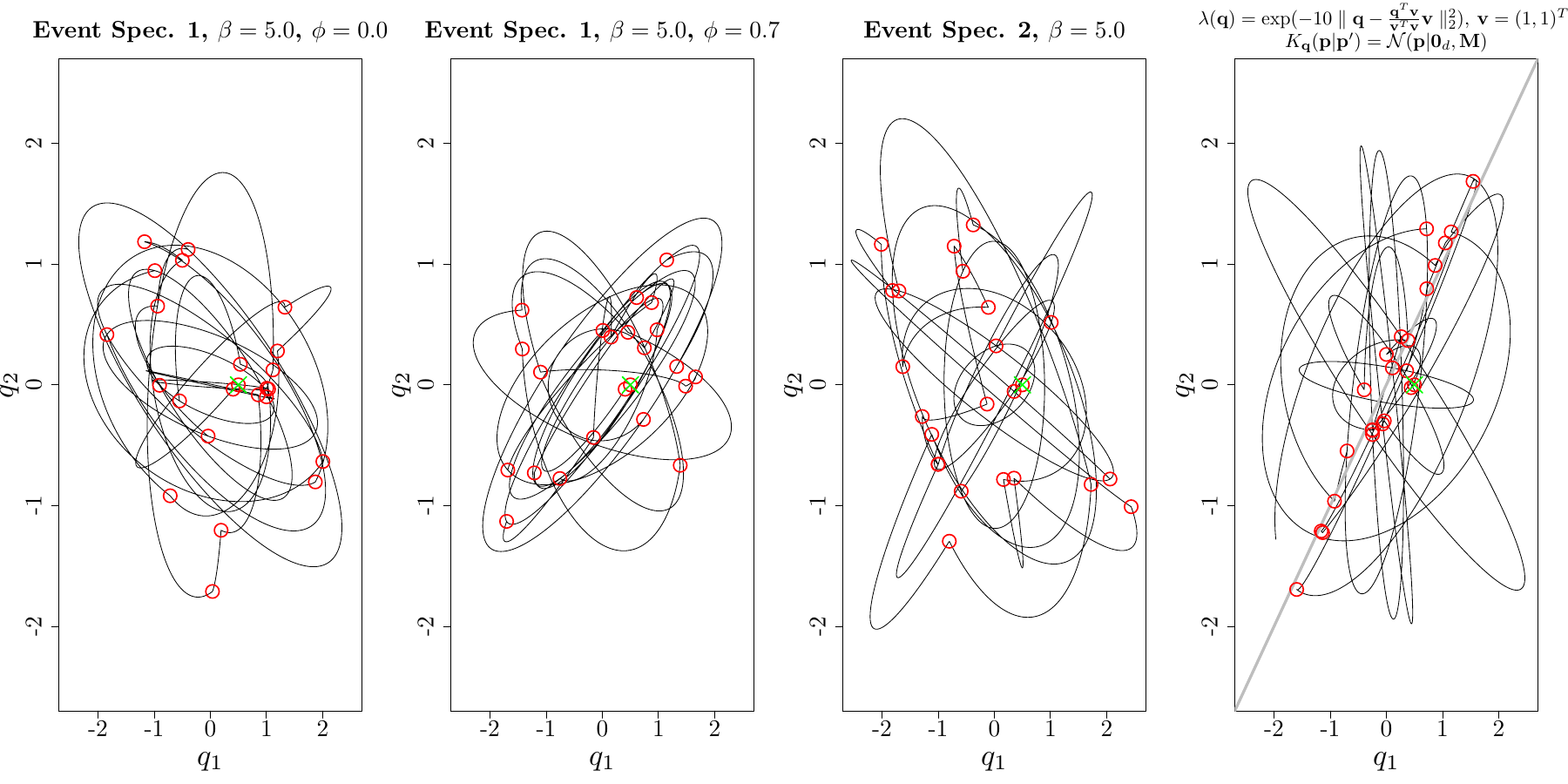}\caption{\label{fig:Examples-of-continuous-trajectories}Examples of $\mathbf{q}$-coordinate
of continuous time HMC trajectories with different event specifications
for a bivariate standard Gaussian target distribution $\pi(\mathbf{q})$.
In all cases, $\mathbf{M}=\mathbf{I}_{2}$ and the shown trajectories
correspond 100 units of time $t$. Events are indicated with red circles,
and the common initial $\mathbf{q}$-coordinate is indicated with
a cross. In the rightmost panel, an event rate favoring events when
the distance between the $\mathbf{q}$-coordinate and its projection
onto the subspace spanned by $\mathbf{v}=(1,1)$ (indicated by a gray
line) is small. }
\end{figure}
Figure \ref{fig:Examples-of-continuous-trajectories} shows examples
of the trajectories of the $\mathbf{q}$-coordinate under continuous
time HMC process for different event specifications for a bivariate
standard Gaussian target. It is seen that the nature of the trajectories
differs, with event specification 1, $\phi=0.7$, visiting a somewhat
narrower ``range'' of orbits relative to that of event specification
1, $\phi=0$. For event specification 1, there is quite large variation
in the ``how much ground'' each between-event trajectory is covering
(several of the between-event trajectories go through multiple identical
cycles), whereas the arc-lengths have less variation under specification
2. Recall still that arc-lengths are only in expectation equal under
event specification 2 as $u$ appearing in (\ref{eq:integrated_event_intensity})
are exponentially distributed. 

The rightmost panel of Figure \ref{fig:Examples-of-continuous-trajectories}
is included as an illustration of the flexibility afforded by GRHMCs
as defined in Section \ref{subsec:GRHMC-as-PDMPs}. An event-rate
favoring events occurring when the $\mathbf{q}$-coordinate is close
to (somewhere on) the subspace defined by $\mathbf{q}_{1}=\mathbf{q}_{2}$
is considered. This is an example of a process where (unlike RHMC)
the embedded discrete time process obtained by considering only the
configuration at events is clearly off target, whereas the continuous
time GRHMC process still has the desired stationary distribution.
Such processes may be an avenue for obtaining between events dynamics
amounting to (an integer multiple) of approximately half orbits. However,
the selection of event-favoring subspaces for general non-Gaussian
target distributions requires further research. From now on, only
event specifications 1, $\phi=0$ and 2 are considered. 

\subsubsection{Moment estimation and ``super-convergence''}

\begin{figure}
\begin{centering}
\includegraphics[scale=0.5]{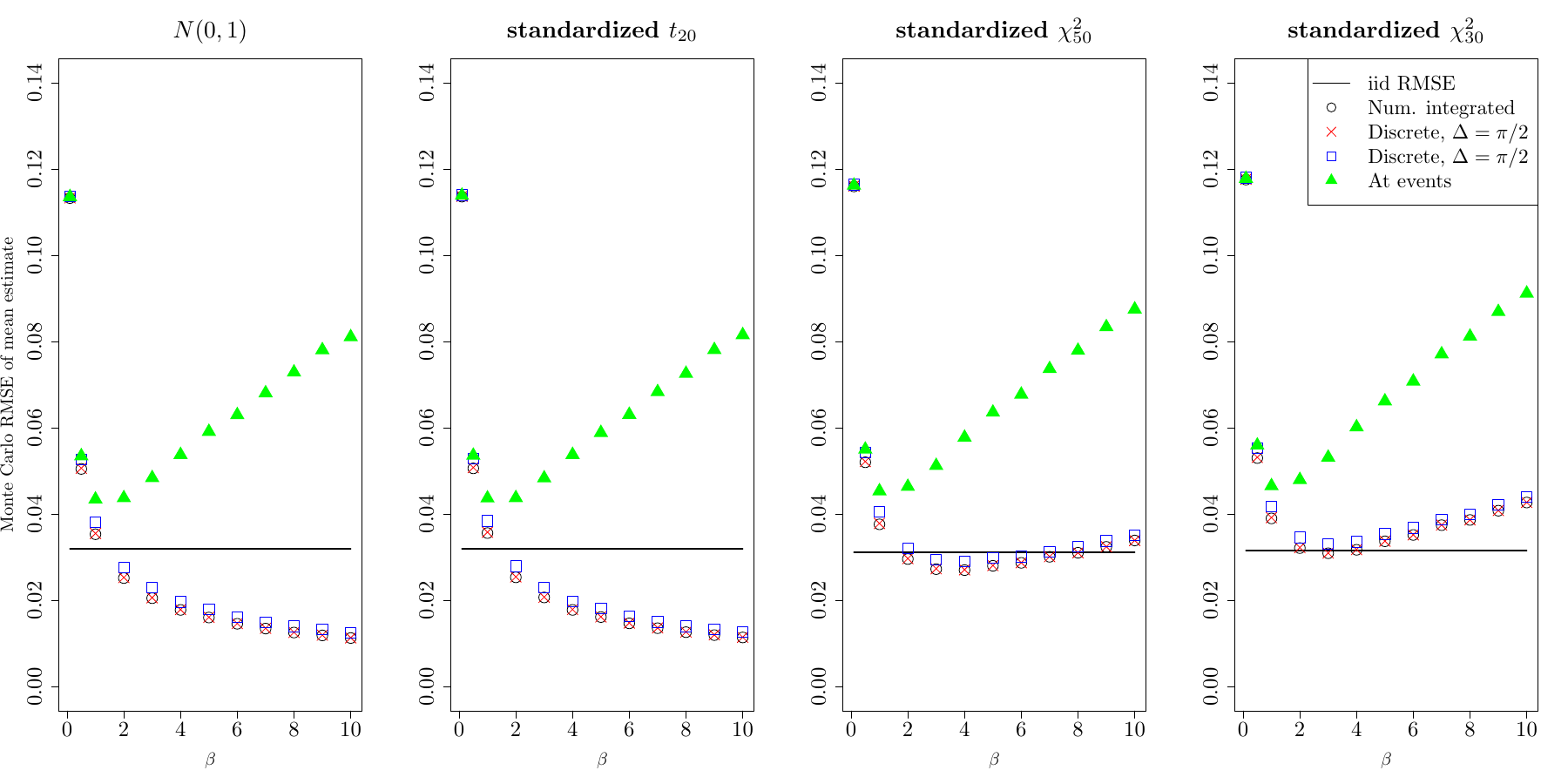}\caption{\label{fig:RMSE-of-estimates}RMSE of estimates of $E(\mathbf{q}_{1})$
from NGRHMC processes using event specification 1 for different values
of the mean inter-event time parameter $\beta$. The panels correspond
to the different univariate target distributions $\pi(\mathbf{q}_{1})$,
which all have zero mean and unit variance. Unit mass matrix $M=1$
was used. The estimates of the mean are based on trajectories of length
$T=1000\frac{\pi}{2}$, and the RMSE estimates are based on 10000
independent replica for each value of $\beta$. Black circles correspond
continuous sampling (\ref{eq:integrated-trajectory}), red $\times$-s
to 1000 equally spaced samples, blue squares to 500 equally spaced
samples and green triangles to samples recorded at events only. The
horizontal lines give the RMSE of 1000 iid samples. The results are
obtained using the numerical methods described in Section \ref{subsec:Numerical-method-for}
with $tol_{a}=tol_{r}=0.001$.}
\par\end{centering}
\end{figure}
To gain some initial insight into the behavior of moment estimation
based on NGRHMC processes, 10000 trajectories of were generated for
4 zero mean, unit variance univariate targets $\pi(\mathbf{q}_{1})$.
Each trajectory was of (time) length $T=1000\frac{\pi}{2}$ preceded
by an equal length of warmup. Event specification 1, $\phi=0$, was
used in all cases, and experiments were repeated for different values
of the inter-event mean time $\beta$ (see Table \ref{tab:The-event-specifications}).
Further, several different sampling strategies were applied to all
produced trajectories. Root mean squared errors (RMSEs) of the $E(\mathbf{q}_{1})$-estimates
are presented in Figure \ref{fig:RMSE-of-estimates}. 

For the standard Gaussian distribution, exactly iid samples obtains
when choosing trajectories of (time) length $\pi/2$ in the hypothetical
HMC method (with exact dynamics, see Section \ref{subsec:Hamiltonian-Monte-Carlo}).
Thus, the (time) lengths of generated Hamiltonian flow (and thus essentially
the computational cost) of NGRHMC and for 1000 iid-samples-producing
hypothetical HMC transitions are the same. As a reference to the NGRHMC
results, the RMSEs based on 1000 iid samples are indicated as horizontal
lines in the plots. For the non-Gaussian targets, the cost of obtaining
RMSEs corresponding to 1000 iid samples using HMC are likely somewhat
higher, and thus the benchmarks are likely somewhat favoring HMC in
a computational cost perspective in these cases.

In all cases, low values of $\beta$ (frequent events) result in poor
results, as the continuous time process approaches the Langevin limit
(\ref{eq:langevin_limit}). The most striking feature of the plot
is that for continuous (black circles) or high frequency sampling
(red $\times$) of the trajectories, RMSEs for the symmetric targets
($N(0,1)$, standardized $t_{20}$) decreases monotonically in $\beta$,
and for the highest $\beta=10.0$ considered is only around 35 percent
of the benchmark in the $N(0,1)$ case. In the univariate Gaussian
target case, this behavior obtains as the between-event Hamiltonian
dynamics, $\mathbf{q}_{1}(t)$, averaged over time, i.e. $\frac{1}{t}\int_{0}^{t}\mathbf{q}_{1}(s)ds$,
converges to the mean of the target as $t\rightarrow\infty$, \emph{regardless
of the initial configuration} $\mathbf{z}(0)$ (see below and Appendix
\ref{sec:Temporal-averages-of}). Thus, in the Gaussian case, momentum
refreshes are not necessary for unbiased estimation of $E(\mathbf{q}_{1})$.
It appears this is also the case for the standardized $t_{20}$-distribution,
but this has not been proved formally so far. 

For the non-symmetric targets, standardized $\chi_{50}^{2}$ and standardized
$\chi_{30}^{2}$, such monotonous behavior is not seen as momentum
refreshes are certainly necessary to obtain high quality estimates.
Too infrequent refreshes (i.e., high $\beta$) result in higher variance
from exploring too few energy level sets. For intermediate values
of $\beta$, the continuous- or high frequency sampling estimates
are still better than or on par with the iid benchmark, where the
edge is lost towards more skewness in the target distribution.

As mentioned in Section \ref{subsec:Piecewise-deterministic-processe},
the continuous time trajectories can either be sampled at discrete
times (\ref{eq:discrete_samples}) or continuously (in practice integrated
numerically within the ODE solver, see below for details) over time
(\ref{eq:integrated-trajectory}), where the former may be thought
of as a crude quadrature approximation to the latter. From Figure
\ref{fig:RMSE-of-estimates}, it is evident that there is little difference
in the high frequency ($\Delta=\pi/2$) discrete time sampling and
continuous sampling. The efficiency deteriorates somewhat with more
infrequent ($\Delta=\pi$) discrete sampling (blue $\boxempty$).
As will be clear in the next section, continuous estimates (\ref{eq:integrated-trajectory})
require minimal additional numerical effort, and it seems advisable
always to use these for moment calculations, whereas rather frequent
discrete samples should be used for other tasks. 

The green triangles represent results obtained when sampling the process
only at event times using the same amount of Hamiltonian trajectory.
It is seen that this practice, which does not exploit the ``between-events''
trajectories, generally lead to inferior results. The exception is
in the random walk-like domain, where frequent events (and thus sampling)
occur, but in this case the underlying process only slowly explores
the target distribution.

\subsection{Choosing event intensities\label{subsec:Choosing-event-intensities}}

The time average property under the univariate Gaussian, illustrated
in the left panel of Figure \ref{fig:RMSE-of-estimates} generalizes
to multivariate Gaussian targets $\pi(\mathbf{q})=\mathcal{N}(\mathbf{q}|\mu,\Sigma)$
as well. Namely, it can be shown (see Appendix \ref{sec:Temporal-averages-of})
that the between-events dynamics $\mathbf{q}(\tau)$ admit unbiased
estimation of $\mu$ without momentum refreshes, i.e. 
\begin{equation}
\frac{1}{T}\int_{0}^{T}\mathbf{\mathbf{C}\mathbf{q}}(\tau)d\tau\underset{T\rightarrow\infty}{\longrightarrow}\mathbf{C}\mu,\;\mathbf{C}\in\mathbb{R}^{p\times d},\label{eq:Gaussian_mean_asymp}
\end{equation}
\emph{regardless of the initial configuration} $\mathbf{z}(0)$. 

Of course, the Gaussian case is not particularly interesting per se.
However, one would presume that for near Gaussian target distributions
(which is frequently the case in Bayesian analysis applications due
to Bernstein-Von Mises effects), the left-hand side (\ref{eq:Gaussian_mean_asymp})
would have only a small variation in $\mathbf{z}(0)$ for large $T$.
Hence such situations would benefit from quite low event intensities/long
durations between events and would allow for very small Monte Carlo
variations in moment estimates akin to those shown in the two leftmost
panels of Figure \ref{fig:RMSE-of-estimates} even in high-dimensional
applications. 

Still, the fast convergence results above are restricted to certain
moments of certain target distributions. It is instructive (and sobering)
to look at the estimation of the second order moment of a univariate
standard Gaussian target distribution (with $\mathbf{M}=1$). In this
case, 
\[
\frac{1}{T}\int_{0}^{T}\mathbf{q}_{1}^{2}(\tau)d\tau\underset{T\rightarrow\infty}{\longrightarrow}\frac{1}{2}\left(\mathbf{q}_{1}^{2}(0)+\mathbf{p}_{1}^{2}(0)\right),
\]
i.e. the dependence on the initial configuration $\mathbf{z}(0)$
does not vanish as the time between events grows, and the second order
moment cannot be estimated reliably without momentum refreshes. 

For a fixed budget of Hamiltonian trajectories, a non-Gaussian target
and/or a non-linear moment, say $E(g(\mathbf{q}))$, and the event
rate tradeoff will have at the endpoints:
\begin{itemize}
\item For ``large $\beta$'', variation in the GRHMC moment estimate is
mainly due to variation between energy level sets, i.e. the variance
of $\lim_{T\rightarrow\infty}\int_{0}^{T}g(\mathbf{q}(t))dt$ as a
function of the initial configuration $\mathbf{z}(0)$. 
\item For ``small $\beta$'', variation in the GRHMC moment estimate comes
mainly from that the underlying process $\mathcal{Z}_{t}$ reverts
to a random walk-like behavior (Langevin-dynamics for constant event
rate).
\end{itemize}
The location of the optimum between these extremes (see e.g., the
two right-most panels in Figure \ref{fig:RMSE-of-estimates}) inherently
depends both on the target distribution and the collection of moments,
say $E(g_{1}(\mathbf{q})),\dots,E(g_{m}(\mathbf{q}))$, one is interested
in. More automatic choices of event rate specifications will be explored
in the numerical experiments discussed below.

\section{Numerical implementation\label{sec:Numerical-implementation}}

The proposed methodology relies on quite accurate simulation of the
Hamiltonian trajectories and associated functionals of the type (\ref{eq:integrated-trajectory}).
This section summarizes numerical implementation of these quantities
based on Runge-Kutta-Nystöm (RKN) methods \citep[see e.g.][Chapter II.14]{10.5555/153158}.
The reader is referred to \citealt{10.5555/153158} for more background
on general purpose ODE solvers. 

In what follows, $\tau$ is used as the time index of the between-events
Hamiltonian dynamics (as opposed to PDMP process time $t)$, and it
is convention that $\tau$ is reset to zero immediately after each
event. RKN methods are particularly well suited for time-homogenous
second order ODE systems on the form 
\begin{equation}
\ddot{\mathbf{y}}(\tau)=\mathbf{F}(\mathbf{y}(\tau)),\;\mathbf{y}\in\mathbb{R}^{n},\;\mathbf{F}:\mathbb{R}^{n}\mapsto\mathbb{R}^{n},\label{eq:gen_sec_order_ode}
\end{equation}
subject to the initial conditions $\mathbf{y}(0)=\mathbf{y}_{0}$,
$\dot{\mathbf{y}}(0)=\text{\ensuremath{\mathbf{z}}}_{0}$. Notice
that when $\mathbf{F}$ does not depend on $\dot{\mathbf{y}}(\tau)$,
RKN methods are substantially more efficient than applying conventional
Runge Kutta methods to an equivalent coupled system of $2n$ first
order equations, say $\dot{\mathbf{y}}(\tau)=\mathbf{w}(\tau),\;\dot{\mathbf{w}}(\tau)=\mathbf{F}(\mathbf{y}(\tau))$. 

A wide range of numerical methods have been developed specifically
for the dynamics of Hamiltonian systems \citep[see e.g.][]{sanzSerna_Calvo,Leimkuhler:2004}.
Such methods typically conserve the symplectic- and time-reversible
properties of the true dynamics, and hence provide reliable long-term
simulations over many (quasi-)orbits. However, for shorter time spans,
typically on the order of up to a few (quasi-)orbits, such symplectic
methods have no edge over conventional methods for second order ODEs
\citep[see e.g.][Section 9.3]{sanzSerna_Calvo}. 

\subsection{Numerical solution of dynamics and functionals\label{subsec:Numerical-method-for}}

In the numerical implementation used in the present work, the between-events
Hamiltonian dynamics are reformulated in terms of the second order
ODE
\begin{equation}
\ddot{\mathbf{q}}(\tau)=\mathbf{M}^{-1}\nabla_{\mathbf{q}}\log(\mathbf{q}(\tau)),\label{eq:second_order_ode}
\end{equation}
which is to be solved for $(\mathbf{q}(\tau),\dot{\mathbf{q}}(\tau))$.
The dynamics of (\ref{eq:second_order_ode}) are equivalent to the
dynamics of (\ref{eq:Ham_eq}) when the initial conditions $(\mathbf{q}(0),\dot{\mathbf{q}}(0)=\mathbf{M}^{-1}\mathbf{p}(0))$
are applied, and the momentum variable for any $\tau$ is recovered
via $\mathbf{p}(\tau)=\mathbf{M}\dot{\mathbf{q}}(\tau)$. 

Further, recall that the proposed methodology relies critically on
the ability to calculate between-events Hamiltonian dynamics functionals
on the form
\begin{equation}
\mathbf{r}_{k}(\tau)=\int_{0}^{\tau}\mathscr{M}_{k}(\mathbf{q}(s))ds,\;k=1,\dots,p,\label{eq:monitoring}
\end{equation}
for a suitably chosen monitoring function $\mathscr{M}:\mathbb{R}^{d}\rightarrow\mathbb{R}^{p}$,
e.g., \emph{for integrated event intensities $\Lambda$} (\ref{eq:integrated_event_intensity})
and continuous sampling (\ref{eq:integrated-trajectory}). To this
end, first observe that $\mathbf{r}(\tau)=\dot{\mathbf{R}}(\tau)$
whenever $\ddot{\mathbf{R}}(\tau)=\mathscr{M}(\mathbf{q}(\tau)),$
with initial conditions $\;\mathbf{R}(0)=\mathbf{0}_{p},\;\dot{\mathbf{R}}(0)=\mathbf{0}_{p}$.
Hence by augmenting (\ref{eq:second_order_ode}) with the monitoring
function, i.e. 
\begin{equation}
\left[\begin{array}{c}
\ddot{\mathbf{q}}(\tau)\\
\ddot{\mathbf{R}}(\tau)
\end{array}\right]=\left[\begin{array}{c}
\mathbf{M}^{-1}\nabla_{\mathbf{q}}\log(\mathbf{q}(\tau))\\
\mathscr{M}(\mathbf{q}(\tau))
\end{array}\right],\label{eq:augmented_sec_order_ode}
\end{equation}
a system on the form (\ref{eq:gen_sec_order_ode}) is obtained. When
solved numerically, (\ref{eq:augmented_sec_order_ode}) produces solutions
both for the dynamics (\ref{eq:Ham_eq} or \ref{eq:second_order_ode})
and the dynamics functional (\ref{eq:monitoring}). Implemented in
this manner, the adaptive step size methodology (discussed in Appendix
\ref{sec:Details-related-to-implementation}) controls both the numerical
error in the Hamiltonian dynamics and the functionals concurrently
(in contrast to \citet{10.1214/19-BA1171} where integrator step sizes
are kept fixed and intermediate integrator steps are included in averages
based on an accept/reject mechanism). 

In this work, the 6th order explicit embedded pair RKN method RKN6(4)6FD
of \citet[Table 2]{DORMAND1987937} was used to solve (\ref{eq:augmented_sec_order_ode}).
Each step of RKN6(4)6FD requires 5 evaluations of the right-hand side
of (\ref{eq:augmented_sec_order_ode}), but of course being a higher
order method, the step sizes may typically be substantially larger
than the stability limit of e.g., the leap frog method. Since the
solution to $\mathbf{R}(\tau)$ is not required per se, trivial modifications
of the mentioned RKN method were done so that it solves only for $\mathbf{s}(\tau)=(\mathbf{q}(\tau),\dot{\mathbf{q}}(\tau),\mathbf{r}(\tau))$.
Further details, and a full algorithm may be found in Section \ref{sec:Details-related-to-implementation}
in the Appendix. It is also worth noticing that simpler, but less
efficient variants of the above algorithm may be written in high level
languages with access to off-the-shelf ODE solvers. Section \ref{sec:Simple-R-implementation}
in the Appendix gives an example written in R.

\subsection{Do numerical errors influence results?\label{subsec:Do-numerical-errors}}

\begin{figure}
\centering{}\includegraphics[scale=0.45]{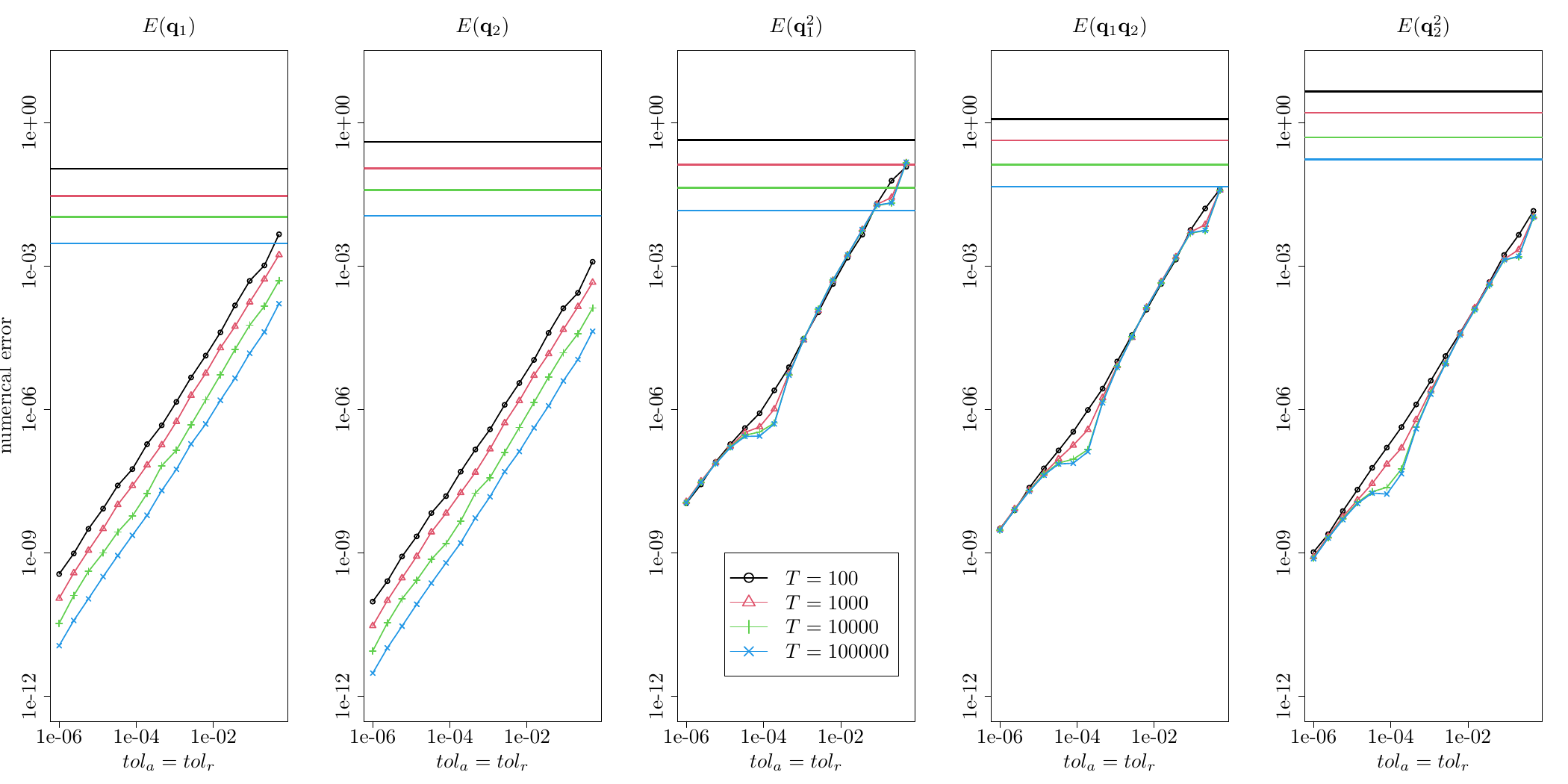}\caption{\label{fig:Numerical-errors-associated}Numerical errors incurred
by RKN integration on the estimation the first and second order moments
of a bivariate Gaussian target distribution. The numerical errors
are relative to a NGRHMC trajectory using the same random numbers
but with exact Hamiltonian dynamics. Both exact and numerically integrated
results are based on continuous sampling. The horizontal axis gives
the error tolerances (in all cases with $tol_{a}=tol_{r}$) applied
in the numerical integrator, and both horizontal and vertical axis
are logarithmic. Dotted lines indicate the root mean squared errors
associated with estimating the indicated moments across many exact
NGRHMC trajectories of different length $T$.}
\end{figure}
To assess how the application of (un-corrected) RKN numerical integrators
for the Hamiltonian dynamics influences overall Monte Carlo estimation,
a small simulation experiment was performed. Specifically, a $N(\mathbf{0}_{2},\Sigma)$
target distribution with 
\begin{equation}
\Sigma=\left[\begin{array}{cc}
1 & 2\\
2 & 8
\end{array}\right],\;\mathbf{M}=\mathbf{I}_{2},\;\lambda=\frac{1}{10}\;\text{and }Q(\mathbf{p}|\mathbf{z}^{\prime})=\mathcal{N}(\mathbf{p}|\mathbf{0}_{2},\mathbf{M}),\label{eq:precision_example_spec}
\end{equation}
was used. Due to the Gaussian nature of the target distribution, the
Hamiltonian dynamics are available in closed form, and hence allow
the comparison with the numerically integrated counterparts. RHMC
trajectories based both on exact and numerically integrated dynamics
were used to estimate the mean and raw second order moments of the
target using continuous sampling. The same initial configuration $\mathcal{Z}(0)$
and the same random numbers were used so that errors in the estimators
based on numerical integration are due only to RKN integration. Figure
\ref{fig:Numerical-errors-associated} shows the RMSEs between estimates
from numerically integrated- and exact RHMC trajectory for different
values of $T$ and the absolute (relative) RKN integrator error tolerance
$tol_{a}$ ($tol_{r}$) (see Appendix \ref{sec:Details-related-to-implementation}
for details). All results are based on independent 50 replications.
Also indicated in the plots as horizontal lines are the RMSEs associated
with estimating the said moments (across multiple independent trajectories)
based on RHMC with exact dynamics.

From Figure \ref{fig:Numerical-errors-associated}, it is seen that
except for very large values of $tol_{a}=tol_{r}$, the numerical
errors are very small relative to the exact estimator RMSEs. From
the plots, absolute and relative error tolerances of around 0.001
appear to be more than sufficient for this case. Interestingly, it
is seen that there is no apparent buildup of numerical errors in the
longer trajectories, suggesting that the incurred errors are not systematically
accumulating and biasing the estimation in any direction. Of course,
this limited experiment does not rule out such biasing behavior in
general. However, the overall the finding here indicate that quite
lax error tolerances are sufficient to make the numerical errors be
negligible relative to overall Monte Carlo variation.

\subsection{Automatic selection of tuning parameters}

A key aim of developing NGRHMC processes is to enable the implementation
of an easy to use and general-purpose code. For this purpose, automatic
selection of tuning parameters is important. This Section describes
the routines for tuning the mass matrix $\mathbf{M}$ and scaling
the event intensity used in the computations described shortly.

\subsubsection{Tuning of mass matrix}

In the present work, only a diagonal mass matrix $\mathbf{M}=\text{diag}(m_{1},\dots,m_{d})$
is considered. Two approaches for choosing each of $m_{1},\dots,m_{d}$
are considered, both exploiting the ability to numerically calculate
temporal averages by augmenting the monitoring function $\mathscr{M}$. 

In the former approach which will be referred to as VARI (variance,
integrated), $m_{i}^{-1}$ is simply set equal to the temporal average
estimate of $Var(\mathbf{q}_{i})$, i.e.
\[
\int_{0}^{t^{*}}\mathbf{\mathcal{Q}}_{i}^{2}(s)ds-\left[\int_{0}^{t^{*}}\mathcal{Q}_{i}(s)ds\right]^{2},
\]
at every event time $t^{*}$ during the warmup period.

In cases where the marginal variances are less informative with respect
to the local scaling of the target distribution, e.g., in the presence
of strong non-linearities or multimodality, a second approach referred
to as ISG (integrated squared gradients) may also pursued. Here overarching
idea for choosing each of $m_{1},\dots,m_{d}$ is to make the square
of each element of the right-hand side of (\ref{eq:second_order_ode}),
averaged over \emph{each integrator step,} in expectation over all
integrator steps, to be equal to 1. This approach is mainly motivated
out of numerical efficiency considerations, where regions of the target
distribution requiring many steps (with short step sizes due to strong
forces $\nabla_{\mathbf{q}}\log\tilde{\pi}(\mathbf{q})$) are disproportionally
weighted when choosing the mass matrix. 

More explicitly, let the $j$th integrator step (during the warmup
period) be originating at time $\tau_{j}$ and have time step size
$\varepsilon_{j}$. Then the mass matrix diagonal $m_{i}$ is taken
to be an exponential moving average (over $j$) of 
\[
\frac{1}{\varepsilon_{j}}\int_{\tau_{j}}^{\tau_{j}+\varepsilon_{j}}\left[\nabla_{\mathbf{q}}\log\tilde{\pi}(\mathbf{q}(s))\right]_{i}^{2}ds.
\]
Notice that the integrated squared gradients are available at negligible
additional cost by augmenting $\mathscr{M}$ in the ODE system (\ref{eq:augmented_sec_order_ode})
with moment functions $\left[\nabla_{\mathbf{q}}\log\tilde{\pi}(\mathbf{q})\right]_{i}^{2},\;i=1,\dots,d$.
Further notice that for a $N(\mu,\mathbf{P}^{-1})$ target distribution,
where $E_{\pi}(\left[\nabla_{\mathbf{q}}\log\tilde{\pi}(\mathbf{q})\right]\left[\nabla_{\mathbf{q}}\log\tilde{\pi}(\mathbf{q})\right]^{T})=\mathbf{P}$,
this approach may (modulus variability in integrator step size) be
seen as a way to \emph{directly estimate the precision matrix diagonal
elements}.

\subsubsection{Tuning of event rates\label{subsec:Tuning-of-event-rates}}

The methodology for tuning the event rates relies of the following
representation of a general event rate $\lambda$:
\[
\lambda(\mathbf{q},\mathbf{p})=\frac{1}{\gamma\beta}\bar{\lambda}(\mathbf{q},\mathbf{p}),\;\gamma>0,\;\beta>0,
\]
where $\bar{\lambda}$ is a ``base line'' event rate (e.g., $\bar{\lambda}=1$
for event specification 1 and 2, and $\bar{\lambda}=\sqrt{\mathbf{p}^{T}\mathbf{M}^{-1}\mathbf{p}}$
for event specification 2). Here $\gamma$ is a user-given scale factor,
say in the range 1 to 20, chosen in the higher range if one expects
good performance with infrequent moment refreshes (see Section \ref{subsec:Choosing-event-intensities}).
Finally, $\beta$ is tuned automatically to reflect each particular
target distribution and event specification. 

Suppose $\iota(\mathbf{z})$ is the distribution the state immediately
after events (which is equal to $\rho(\mathbf{z})$ for RHMC but may
also differ substantially relative to $\rho(\mathbf{z})$ as seen
in the rightmost panel of Figure \ref{fig:Examples-of-continuous-trajectories}).
The objective of the automatic event rate tuning is given by 
\begin{equation}
\underset{\mathbf{z}(0)\sim\iota(\mathbf{z})}{E}\beta^{-1}\Upsilon(\mathbf{z}(0))=1,\text{ where }\Upsilon(\mathbf{z}(0))=\int_{0}^{\omega(\mathbf{z}(0))}\bar{\lambda}(\mathbf{q}(\tau),\mathbf{p}(\tau))d\tau,\label{eq:adapt_beta_target}
\end{equation}
and where $\omega$ is the ``U-turn'' time \citep{JMLR:v15:hoffman14a}
\[
\omega(\mathbf{z}(0))=\inf\left\{ \tau>0\;:\;(\mathbf{q}(\tau)-\mathbf{q}(0))^{T}\mathbf{p}(\tau)<0\right\} ,
\]
of the dynamics (\ref{eq:Ham_eq}) initialized at $\mathbf{z}(0)$
\citep[See also][for a similar development]{1810.04449}. The rationale
behind (\ref{eq:adapt_beta_target}) is that (for $\gamma=1$) the
expected integrated event rate $\Lambda$ (see Equation \ref{eq:integrated_event_intensity})
evaluated at corresponding U-turn time is equal to $E(u)$. 

In the present implementation, $\Upsilon(\mathbf{z}(0))$ is computed
for each event during warmup with $\mathbf{z}(0)$ being the state
immediately after the events. Subsequently, $\beta^{-1}$ is updated
(also during the warmup period only) at each event according to an
exponential moving average over the already computed $\Upsilon$s.
The exponential moving average is used as the older realizations of
$\Upsilon$ are typically recorded with a different mass matrix encountered
earlier in the mass matrix adaptation process. Computing $\Upsilon(\mathbf{z}(0))$
for each event incurs only modest additional costs, since $\Upsilon$
is a scalar integrated quantity computed over Hamiltonian dynamics
that are integrated numerically anyway. However, if the next event
occurs before the U-turn time $\omega$, further integration steps
are performed until $\omega$ is reached. The additional (post-event)
Hamiltonian dynamics used to locate $\omega$ are subsequently discarded. 

\section{Numerical experiments\label{sec:Numerical-experiments}}

This section considers numerical experiments and benchmarking of the
proposed method against the NUTS-HMC implementation in Stan (rstan
version 2.21.2). Like Stan, the proposed methodology has been implemented
as an R package (pdphmc) with main computational tasks done in C++,
and relies, like rstan, on the Stan Math Library \citep{JSSv076i01}
for automatic differentiation and probability- and linear algebra
computations. 

All computations in this section were carried out on a 2020 Macbook
pro with a 2.6 GHz Intel Core i7 processor, under R version 4.0.3.
In line with the findings in Section \ref{subsec:Do-numerical-errors},
the default integrator tolerances $tol_{a}=tol_{r}=0.001$ are used
for pdphmc unless otherwise noted. The package pdphmc, and code and
data for reproducing the reported results is available at \href{https://github.com/torekleppe/PDPHMCpaperCode}{https://github.com/torekleppe/PDPHMCpaperCode}.

In order to compare the performance of the methods, their Effective
Sample Size (ESS) \citep{geyer1992} per computing time \citep[see e.g.][]{girolami_calderhead_11}
is taken as the main statistic. Consider a sample dependent of dependent
random variables $\eta_{i},\;i=1,\dots,N$, each having the same marginal
distribution. The ESS gives the number of hypothetical iid samples
(with distribution equal to that of $\eta_{1}$) required to obtain
a mean estimator with the same variance as $N^{-1}\sum_{i=1}^{N}\eta_{i}$
. An ESS-based approach is taken also here, but in order to obtain
ESSes for moments estimated by for integrated quantities (\ref{eq:integrated-trajectory}),
the following approach was taken: For a given number of samples, say
$N$, rewrite the left-hand side of (\ref{eq:integrated-trajectory})
as
\begin{equation}
\frac{1}{T}\int_{0}^{T}g(\mathcal{Z}(t))dt=\frac{1}{N}\sum_{i=1}^{N}\eta_{i},\;\text{where }\eta_{i}=\Delta^{-1}\int_{(i-1)\Delta}^{i\Delta}g(\mathcal{Z}(t))dt,\;\Delta=\frac{T}{N}.\label{eq:int_ESS_subdiv}
\end{equation}
Let $\widehat{ESS}_{i}(\eta_{i})$ denote an estimator of the ESS
of dependent sample $\eta_{i}$. Then 
\begin{equation}
\frac{\widehat{Var}_{i}(g(\mathcal{Z}(\Delta i)))}{\widehat{Var}_{i}(\eta_{i})}\widehat{ESS}_{i}(\eta_{i})\label{eq:int_ESS}
\end{equation}
is taken to be an estimator of ESS represented by moment estimator
$T^{-1}\int_{0}^{T}g(\mathcal{Z}(t))dt$, expressed in terms of iid
samples of $g(\mathbf{q})$. Equation \ref{eq:int_ESS} takes into
account both that $\widehat{Var}_{i}(\eta_{i})$ tends to be smaller
than $\widehat{Var}_{i}(g(\mathcal{Z}(\Delta i)))$ due to the temporal
averaging in (\ref{eq:int_ESS_subdiv}), but on the other hand $\eta_{i}$
tends to exhibit a stronger autocorrelation than discrete time samples
$g(\mathcal{Z}(\Delta i))$. Throughout this text, the ESS estimation
procedure in rstan (see R-function rstan::monitor(), output ``n\_eff'')
was used for estimating ESS from samples. In addition, the largest
(over sampled quantities) Gelman-Rubin $\hat{R}$ statistics ($\max\hat{R}$)
\citep{GelmanBDA3} were computed using the same function. For pdphmc,
the reported $\max\hat{R}$ are computed for the discretely sampled
processes.

Comparing the performance of different MCMC methods is intrinsically
hard. Care has been taken so that all code is written in the same
language and compiled with the same compiler on the same computer
and so on. Still, in the present context one must also consider that
rstan is based on an ``exact'' MCMC scheme whereas pdphmc will in
general be subject to (arbitrarily small, at the cost of more computing,)
biases stemming from the use of uncorrected numerical integration.
On the other hand, as demonstrated e.g., in Figure \ref{fig:Initial-experiment-funnel},
chains of finite length generated by rstan may fail to reflect the
target distribution in cases where no visible bias is exhibited by
pdphmc due to the adaptive nature of the applied integrators. The
relative weighting of these features naturally depends on the application
at hand, and therefore preclude strong conclusions regarding the relative
performance of the methods.

In what follows, three numerical experiments are presented. A further
experiment, based on a crossed random effects model for the Salamander
data is described in Section \ref{sec:The-salamander-mating} of the
Appendix. For the Salamander data, pdphmc is found to be on par or
somewhat more efficient than rstan.

\subsection{Funnel distribution\label{subsec:Funnel-distribution}}

The Funnel distribution $\mathbf{q}_{1}\sim N(0,1)$, $\mathbf{q}_{2}|\mathbf{q}_{1}\sim N(0,\exp(3\mathbf{q}_{1}))$
\citep[funnel distributions may be traced back to][]{neal2003}, constituting
the first numerical example, has already been encountered in Section
\ref{subsec:Hamiltonian-Monte-Carlo} and Figure \ref{fig:Initial-experiment-funnel}.
This very simple example may be considered as a ``model problem''
displaying similar behavior as for targets associated with Bayesian
hierarchical models (where $q_{1}$ plays the role of latent field
log-scale parameter, and $q_{2}$ plays the role of the latent field
it self). 

For both rstan and pdphmc, 10 independent chains/trajectories were
run with identity mass matrices. For rstan, each of these chains had
10,000 transitions with 5,000 discarded as warmup. The number of warmup
iterations is larger than the default 1000 to allow for best possible
integrator step size adaptation. The remaining tuning parameters of
rstan are the default. Note that rstan outputs a substantial number
of warnings related to diverged transitions for all values of $\delta$. 

For pdphmc, the trajectories were of length $T=100,000$, sampled
discretely $N=10,000$ times and with the former half of samples discarded
as warmup. For such high sampling frequency, continuous samples yield
similar results as the discrete samples, and are not discussed further
here. A constant event rate $\lambda=\beta^{-1}$ was applied, and
$\beta$ was adapted with scale factor $\gamma=2$ using the methodology
described in Section \ref{subsec:Tuning-of-event-rates}. The adaptive
selection resulted in values of $\beta$ between 2.1 and 4.7 across
the 10 trajectories, which again translates to between 0.21 and 0.48
discrete time samples per (between-events) Hamiltonian trajectory. 

It has already been confirmed visually from Figure \ref{fig:Initial-experiment-funnel}
that the output of rstan does not fully explore the target distribution
as fixed time step size integration is broadly speaking unsuitable
for this problem. Consequently, ESSes for rstan are not presented.
pdphmc produces around 800 effective samples per second for the log-scale
parameter $\mathbf{q}_{1}$. This is close to double what one obtains
by calculating time-weighted ESS for the (still defective) $\delta=0.999$
rstan chains, indicating the the proposed methodology is highly competitive
for difficult problems (as even smaller fixed time steps would be
required to obtain proper convergence). Further, the default integrator
tolerances $tol_{a}=tol_{r}=0.001$ lead to biases (relative to the
theoretical process) that are not detectable from the right panel
of Figure \ref{fig:Initial-experiment-funnel}.

\subsection{Smile-shaped distribution}

\begin{table}
\centering{}%
\begin{tabular}{ccccccccccccccc}
\hline 
{\footnotesize{}$\gamma$} & {\footnotesize{}sampling} &  & \multicolumn{2}{c}{{\footnotesize{}$\mathbf{q}_{1}$}} &  & \multicolumn{2}{c}{{\footnotesize{}$\underset{k\in\{2,11\}}{\min}ESS(\mathbf{q}_{k})$}} &  & \multicolumn{2}{c}{{\footnotesize{}$\underset{k\in\{2,11\}}{\max}ESS(\mathbf{q}_{k})$}} &  & {\footnotesize{}$E(\mathbf{q}_{1})$} & {\footnotesize{}$E(\mathbf{q}_{2})$} & {\footnotesize{}CPU time }\tabularnewline
\cline{4-5} \cline{5-5} \cline{7-8} \cline{8-8} \cline{10-11} \cline{11-11} 
 &  &  & {\footnotesize{}ESS} & {\footnotesize{}$\frac{\text{ESS}}{\text{CPU time}}$} &  & {\footnotesize{}ESS} & {\footnotesize{}$\frac{\text{ESS}}{\text{CPU time}}$} &  & {\footnotesize{}ESS} & {\footnotesize{}$\frac{\text{ESS}}{\text{CPU time}}$} &  & {\footnotesize{}(exact $=0$)} & {\footnotesize{}(exact $=1$)} & {\footnotesize{}(s)}\tabularnewline
\hline 
\multicolumn{15}{c}{{\footnotesize{}rstan ($\max\hat{R}=1.247$)}}\tabularnewline
\hline 
 &  &  & {\footnotesize{}19 } & {\footnotesize{}10} &  & {\footnotesize{}29} & {\footnotesize{}15} &  & {\footnotesize{}32} & {\footnotesize{}17} &  & {\footnotesize{}-0.15 } & {\footnotesize{}1.04 } & {\footnotesize{}1.9 }\tabularnewline
\hline 
\multicolumn{15}{c}{{\footnotesize{}pdphmc, event specification 1, $\phi=0$ ($\gamma=2$
: $\max\hat{R}=1.006$, $\gamma=10$ : $\max\hat{R}=1.015$)}}\tabularnewline
\hline 
{\footnotesize{}2} & {\footnotesize{}D} &  & {\footnotesize{}1323} & {\footnotesize{}779} &  & {\footnotesize{}1065} & {\footnotesize{}627} &  & {\footnotesize{}1215} & {\footnotesize{}716} &  & {\footnotesize{}-0.02 } & {\footnotesize{}1.03 } & {\footnotesize{}1.7 }\tabularnewline
{\footnotesize{}2} & {\footnotesize{}C} &  & {\footnotesize{}1319} & {\footnotesize{}777} &  & {\footnotesize{}1115} & {\footnotesize{}657} &  & {\footnotesize{}1150} & {\footnotesize{}678} &  & {\footnotesize{}-0.02 } & {\footnotesize{}1.02 } & \tabularnewline
{\footnotesize{}10} & {\footnotesize{}D} &  & {\footnotesize{}2213} & {\footnotesize{}1304} &  & {\footnotesize{}608} & {\footnotesize{}358} &  & {\footnotesize{}634} & {\footnotesize{}374} &  & {\footnotesize{}0.01 } & {\footnotesize{}1.00 } & {\footnotesize{}1.7 }\tabularnewline
{\footnotesize{}10} & {\footnotesize{}C} &  & {\footnotesize{}2233} & {\footnotesize{}1316} &  & {\footnotesize{}613} & {\footnotesize{}361} &  & {\footnotesize{}630} & {\footnotesize{}371} &  & {\footnotesize{}0.01 } & {\footnotesize{}1.01 } & \tabularnewline
\hline 
\multicolumn{15}{c}{{\footnotesize{}pdphmc, event specification 2 ($\gamma=2$ : $\max\hat{R}=1.007$,
$\gamma=10$ : $\max\hat{R}=1.009$)}}\tabularnewline
\hline 
{\footnotesize{}2} & {\footnotesize{}D} &  & {\footnotesize{}1094} & {\footnotesize{}645} &  & {\footnotesize{}1130} & {\footnotesize{}666} &  & {\footnotesize{}1183} & {\footnotesize{}697} &  & {\footnotesize{}-0.02 } & {\footnotesize{}0.98 } & {\footnotesize{}1.7}\tabularnewline
{\footnotesize{}2} & {\footnotesize{}C} &  & {\footnotesize{}1094} & {\footnotesize{}645} &  & {\footnotesize{}1153} & {\footnotesize{}679} &  & {\footnotesize{}1184} & {\footnotesize{}698} &  & {\footnotesize{}-0.02 } & {\footnotesize{}0.97 } & \tabularnewline
{\footnotesize{}10} & {\footnotesize{}D} &  & {\footnotesize{}2276} & {\footnotesize{}1341} &  & {\footnotesize{}920} & {\footnotesize{}542} &  & {\footnotesize{}967} & {\footnotesize{}570} &  & {\footnotesize{}0.01 } & {\footnotesize{}1.03 } & {\footnotesize{}1.7 }\tabularnewline
{\footnotesize{}10} & {\footnotesize{}C} &  & {\footnotesize{}2303 } & {\footnotesize{}1357} &  & {\footnotesize{}927} & {\footnotesize{}546} &  & {\footnotesize{}948} & {\footnotesize{}559} &  & {\footnotesize{}0.01 } & {\footnotesize{}1.03 } & \tabularnewline
\hline 
\end{tabular}\caption{\label{tab:Results-for-the-smile}Results for the ``smile''-shaped
target distribution (\ref{eq:smile_1},\ref{eq:smile_2}). The results
are based on 10 independent replica, and ESSes and ESSes per computing
time (best in bold font) are from the combined results over these
replica. For rstan, each replica consists of the 10,000 transitions,
with the former 5,000 discarded as warmup. For pdphmc, ISG-type mass
matrix, trajectories of length $T=25,000$ divided evenly between
warmup and sampling, and 1000 discrete samples were used. The presented
CPU times are the total time spent by all chains/trajectories during
the post warmup period. For each configuration of pdphmc, results
from both discrete sampling (D) and continuous sampling (C) are presented. }
\end{table}
To further explore the performance of pdphmc applied to a highly non-linear
target distributions; the ``smile''-shaped distribution
\begin{align}
\mathbf{q}_{k}|\mathbf{q}_{1} & \sim N(\mathbf{q}_{1}^{2},0.5^{2}),\;k=2,\dots,11,\label{eq:smile_1}\\
\mathbf{q}_{1} & \sim N(0,1).\label{eq:smile_2}
\end{align}
is considered. The results for various settings of pdphmc and rstan
are given in Table \ref{tab:Results-for-the-smile}. From the Table,
it is seen that rstan has substantial convergence problems with the
largest Gelman-Rubin $\hat{R}>1.05$, whereas the various settings
of pdphmc reliably explores the target. Choosing longer trajectories
($\gamma=10$) results in higher sampling efficiency for the marginally
standard Gaussian $\mathbf{q}_{1}$, whereas for the non-Gaussian
components $\mathbf{q}_{2:11}$, shorter trajectories are more efficient.
Comparing the event specifications 1 and 2, it is seen that none of
them produces uniformly better results. 

\subsection{Logistic regression}

\begin{table}
\begin{centering}
{\footnotesize{}}%
\begin{tabular}{cccccccccccc}
\hline 
{\footnotesize{}$\gamma$} & {\footnotesize{}Sampling} &  & \multicolumn{2}{c}{{\footnotesize{}$\min_{j}$ $\widehat{ESS}(\boldsymbol{\beta}_{j})$}} &  & \multicolumn{2}{c}{{\footnotesize{}$\text{median}_{j}$ $\widehat{ESS}(\boldsymbol{\beta}_{j})$}} &  & \multicolumn{2}{c}{{\footnotesize{}$\max_{j}$ $\widehat{ESS}(\boldsymbol{\beta}_{j})$}} & {\footnotesize{}CPU}\tabularnewline
\cline{4-5} \cline{5-5} \cline{7-8} \cline{8-8} \cline{10-11} \cline{11-11} 
 &  &  & {\footnotesize{}ESS} & {\footnotesize{}$\frac{\text{ESS}}{\text{CPU time}}$} &  & {\footnotesize{}ESS} & {\footnotesize{}$\frac{\text{ESS}}{\text{CPU time}}$} &  & {\footnotesize{}ESS} & {\footnotesize{}$\frac{\text{ESS}}{\text{CPU time}}$} & {\footnotesize{}time}\tabularnewline
\hline 
\multicolumn{12}{c}{{\footnotesize{}rstan ($\max\hat{R}=1.003$)}}\tabularnewline
\hline 
 &  &  & {\footnotesize{}9676} & {\footnotesize{}2125 } &  & {\footnotesize{}13450 } & {\footnotesize{}2953 } &  & {\footnotesize{}15812 } & {\footnotesize{}3472 } & {\footnotesize{}4.55}\tabularnewline
\hline 
\multicolumn{12}{c}{{\footnotesize{}pdphmc, event specification 1 ($\gamma=5$ : $\max\hat{R}=1.011$,
$\gamma=20$ : $\max\hat{R}=1.020$)}}\tabularnewline
\hline 
{\footnotesize{}5} & {\footnotesize{}D} &  & {\footnotesize{}11350} & {\footnotesize{}1425 } &  & {\footnotesize{}23114} & {\footnotesize{}2903 } &  & {\footnotesize{}40000} & {\footnotesize{}5023 } & \multirow{2}{*}{{\footnotesize{}7.96}}\tabularnewline
{\footnotesize{}5} & {\footnotesize{}C} &  & {\footnotesize{}18220} & {\footnotesize{}2288} &  & {\footnotesize{}32967 } & {\footnotesize{}4140 } &  & {\footnotesize{}71079} & \textbf{\footnotesize{}8926} & \tabularnewline
{\footnotesize{}20} & {\footnotesize{}D} &  & {\footnotesize{}11853} & {\footnotesize{}1490} &  & {\footnotesize{}29752 } & {\footnotesize{}3740 } &  & {\footnotesize{}40000 } & {\footnotesize{}5028} & \multirow{2}{*}{{\footnotesize{}7.96 }}\tabularnewline
{\footnotesize{}20} & {\footnotesize{}C} &  & {\footnotesize{}20423} & \textbf{\footnotesize{}2567} &  & {\footnotesize{}36019 } & {\footnotesize{}4528 } &  & {\footnotesize{}69623 } & {\footnotesize{}8752} & \tabularnewline
\hline 
\multicolumn{12}{c}{{\footnotesize{}pdphmc, event specification 2 ($\gamma=5$ : $\max\hat{R}=1.007$,
$\gamma=20$ : $\max\hat{R}=1.027$)}}\tabularnewline
\hline 
{\footnotesize{}5} & {\footnotesize{}D} &  & {\footnotesize{}11281} & {\footnotesize{}1362} &  & {\footnotesize{}23185 } & {\footnotesize{}2799 } &  & {\footnotesize{}40000 } & {\footnotesize{}4829} & \multirow{2}{*}{{\footnotesize{}8.28  }}\tabularnewline
{\footnotesize{}5} & {\footnotesize{}C} &  & {\footnotesize{}18317} & {\footnotesize{}2211} &  & {\footnotesize{}30771 } & {\footnotesize{}3715 } &  & {\footnotesize{}66928 } & {\footnotesize{}8080 } & \tabularnewline
{\footnotesize{}20} & {\footnotesize{}D} &  & {\footnotesize{}12780} & {\footnotesize{}1565} &  & {\footnotesize{}32653 } & {\footnotesize{}3999} &  & {\footnotesize{}40000 } & {\footnotesize{}4899} & \multirow{2}{*}{{\footnotesize{}8.16 }}\tabularnewline
{\footnotesize{}20} & {\footnotesize{}C} &  & {\footnotesize{}20902} & {\footnotesize{}2560} &  & {\footnotesize{}43303 } & \textbf{\footnotesize{}5304} &  & {\footnotesize{}70584 } & {\footnotesize{}8645} & \tabularnewline
\hline 
\end{tabular}\caption{\label{tab:German-credit-data,}ESSes and time weighted ESSes for
the logistic regression model (\ref{eq:logistic_1},\ref{eq:logistic_2})
applied to the German credit data. All figures are based on 10 independent
chains/trajectories. For rstan, the default 1000 warmup transitions
followed by 1000 sampling transitions were used. For pdphmc, a VARI
mass matrix, $T=5000$ divided evenly between warmup and sampling
and 1000 discrete samples per trajectory were used. }
\par\end{centering}
\end{table}
The next example model is a basic logistic regression model
\begin{align}
\mathbf{y}_{i}|\boldsymbol{\beta} & \sim Beroulli(\mathbf{p}_{i}),\;\text{logit}(\mathbf{\mathbf{p}}_{i})=\mathbf{x}_{i,\cdot}^{T}\boldsymbol{\beta},\;i=1,\dots,n,\label{eq:logistic_1}\\
\boldsymbol{\beta} & \sim N(\mathbf{0}_{p},100\mathbf{I}_{p})\label{eq:logistic_2}
\end{align}
applied to the German credit data \citep[see e.g.][]{reason:MicSpiTay94a}
which has $n=1000$ examples and $p=25$ covariates (including a constant
term). This example is included to measure the performance relative
to rstan on an ``easy'' target distribution \citep{chopin2017}. 

Results for pdphmc and rstan are provided in Table \ref{tab:German-credit-data,}.
It is seen that rstan produces less variation in the ESSes across
the different parameters than pdphmc, which presumably is related
to the mass matrix adaptation. The discrete samples cases of pdphmc
have a somewhat slower minimum ESS performance but on par or better
median- and maximum ESS performance. For this target distribution,
continuous samples for the first moment of $\boldsymbol{\beta}|\mathbf{y}$
substantially improves the performance of pdphmc relative to the corresponding
discretely sampled counterparts in all cases.

\section{Dynamic Inverted Wishart model for realized covariances\label{subsec:Dynamic-Inverted-Wishart}}

\subsection{Model and data}

\begin{table}
\begin{centering}
{\footnotesize{}}%
\begin{tabular}{clccccccc}
\hline 
 &  & {\footnotesize{}rstan} &  & \multicolumn{2}{c}{{\footnotesize{}pdphmc, event spec. 1, $\gamma=10.0$}} &  & \multicolumn{2}{c}{{\footnotesize{}pdphmc, event spec. 2, $\gamma=10.0$}}\tabularnewline
 & {\footnotesize{}CPU time (S,A) } & {\footnotesize{}(14605, 72127) } &  & \multicolumn{2}{c}{{\footnotesize{}(10629, 25244) }} &  & \multicolumn{2}{c}{{\footnotesize{}(10189, 24960) }}\tabularnewline
 & {\footnotesize{}$\max\hat{R}$} & {\footnotesize{}1.003} &  & \multicolumn{2}{c}{{\footnotesize{}1.032}} &  & \multicolumn{2}{c}{{\footnotesize{}1.034}}\tabularnewline
\cline{5-6} \cline{6-6} \cline{8-9} \cline{9-9} 
 & {\footnotesize{}sampling} &  &  & {\footnotesize{}D} & {\footnotesize{}C} &  & {\footnotesize{}D} & {\footnotesize{}C}\tabularnewline
{\footnotesize{}$\boldsymbol{\mu}$} & {\footnotesize{}ESS (min, max)} & {\footnotesize{}(14304, 18487) } &  & {\footnotesize{}(12906, 28173) } & {\footnotesize{}(14467, 28370) } &  & {\footnotesize{}(14749, 38387) } & {\footnotesize{}(14224, 41821)}\tabularnewline
 & {\footnotesize{}ESS/S (min, max)} & {\footnotesize{}(0.98, 1.27) } &  & {\footnotesize{}(1.21, 2.65) } & {\footnotesize{}(1.36, 2.67) } &  & {\footnotesize{}(1.45, 3.77) } & {\footnotesize{}(1.40, 4.10)}\tabularnewline
 & {\footnotesize{}ESS/A (min, max)} & {\footnotesize{}(0.20, 0.26) } &  & {\footnotesize{}(0.51, 1.12) } & {\footnotesize{}(0.57, 1.12) } &  & {\footnotesize{}(0.59, 1.54) } & {\footnotesize{}(0.57, 1.68)}\tabularnewline
{\footnotesize{}$\boldsymbol{\sigma}$} & {\footnotesize{}ESS (min, max)} & {\footnotesize{}(14757, 20011) } &  & {\footnotesize{}(28447, 38922) } & {\footnotesize{}(35971, 42891) } &  & {\footnotesize{}(30227, 40000) } & {\footnotesize{}(34283, 44316)}\tabularnewline
 & {\footnotesize{}ESS/S (min, max)} & {\footnotesize{}(1.01, 1.37) } &  & {\footnotesize{}(2.68, 3.66) } & {\footnotesize{}(3.38, 4.04) } &  & {\footnotesize{}(2.97, 3.93) } & {\footnotesize{}(3.36, 4.35)}\tabularnewline
 & {\footnotesize{}ESS/A (min, max)} & {\footnotesize{}(0.20, 0.28) } &  & {\footnotesize{}(1.13, 1.54) } & {\footnotesize{}(1.42, 1.70) } &  & {\footnotesize{}(1.21, 1.60) } & {\footnotesize{}(1.37, 1.78)}\tabularnewline
{\footnotesize{}$\boldsymbol{\delta}$} & {\footnotesize{}ESS (min, max)} & {\footnotesize{}(13624, 16678) } &  & {\footnotesize{}(16145, 28401) } & {\footnotesize{}(18109, 31180) } &  & {\footnotesize{}(15749, 25582) } & {\footnotesize{}(9464, 29271)}\tabularnewline
 & {\footnotesize{}ESS/S (min, max)} & {\footnotesize{}(0.93, 1.14) } &  & {\footnotesize{}(1.52, 2.67) } & {\footnotesize{}(1.70, 2.93) } &  & {\footnotesize{}(1.55, 2.51) } & {\footnotesize{}(0.93, 2.87)}\tabularnewline
 & {\footnotesize{}ESS/A (min, max)} & {\footnotesize{}(0.19, 0.23) } &  & {\footnotesize{}(0.64, 1.13) } & {\footnotesize{}(0.72, 1.24) } &  & {\footnotesize{}(0.63, 1.02) } & {\footnotesize{}(0.38, 1.17)}\tabularnewline
{\footnotesize{}$\mathbf{H}$} & {\footnotesize{}ESS (min, max)} & {\footnotesize{}(12195, 22072) } &  & {\footnotesize{}(35292, 40000) } & {\footnotesize{}(46273, 69861) } &  & {\footnotesize{}(33089, 40000) } & {\footnotesize{}(43658, 70275)}\tabularnewline
 & {\footnotesize{}ESS/S (min, max)} & {\footnotesize{}(0.84, 1.51) } &  & {\footnotesize{}(3.32, 3.76) } & {\footnotesize{}(4.35, 6.57) } &  & {\footnotesize{}(3.25, 3.93) } & {\footnotesize{}(4.28, 6.90)}\tabularnewline
 & {\footnotesize{}ESS/A (min, max)} & {\footnotesize{}(0.17, 0.31) } &  & {\footnotesize{}(1.40, 1.58) } & {\footnotesize{}(1.83, 2.77) } &  & {\footnotesize{}(1.33, 1.60) } & {\footnotesize{}(1.75, 2.82)}\tabularnewline
{\footnotesize{}$\nu$} & {\footnotesize{}ESS } & {\footnotesize{}17797 } &  & {\footnotesize{}40000} & {\footnotesize{}53388} &  & {\footnotesize{}40000 } & {\footnotesize{}56313}\tabularnewline
 & {\footnotesize{}ESS/S } & {\footnotesize{}1.22 } &  & {\footnotesize{}3.76 } & {\footnotesize{}5.02 } &  & {\footnotesize{}3.93 } & {\footnotesize{}5.53}\tabularnewline
 & {\footnotesize{}ESS/A } & {\footnotesize{}0.25} &  & {\footnotesize{}1.58} & {\footnotesize{}2.11 } &  & {\footnotesize{}1.60 } & {\footnotesize{}2.26}\tabularnewline
{\footnotesize{}$\mathbf{z}_{1,\cdot}$} & {\footnotesize{}ESS (min, max)} & {\footnotesize{}(22130, 24317) } &  & {\footnotesize{}(30533, 40000) } & {\footnotesize{}(37065, 70920) } &  & {\footnotesize{}(40000, 40000) } & {\footnotesize{}(64787, 68952)}\tabularnewline
 & {\footnotesize{}ESS/S (min, max)} & {\footnotesize{}(1.52, 1.67) } &  & {\footnotesize{}(2.87, 3.76) } & {\footnotesize{}(3.49, 6.67) } &  & {\footnotesize{}(3.93, 3.93) } & {\footnotesize{}(6.36, 6.77)}\tabularnewline
 & {\footnotesize{}ESS/A (min, max)} & {\footnotesize{}(0.31, 0.34) } &  & {\footnotesize{}(1.21, 1.58) } & {\footnotesize{}(1.47, 2.81) } &  & {\footnotesize{}(1.60, 1.60) } & {\footnotesize{}(2.60, 2.76)}\tabularnewline
{\footnotesize{}$\mathbf{x}_{1,\cdot}$} & {\footnotesize{}ESS (min, max)} & {\footnotesize{}(22149, 24858) } &  & {\footnotesize{}(27525, 40000) } & {\footnotesize{}(33768, 70763) } &  & {\footnotesize{}(40000, 40000) } & {\footnotesize{}(63230, 68873)}\tabularnewline
 & {\footnotesize{}ESS/S (min, max)} & {\footnotesize{}(1.52, 1.70) } &  & {\footnotesize{}(2.59, 3.76) } & {\footnotesize{}(3.18, 6.66) } &  & {\footnotesize{}(3.93, 3.93) } & {\footnotesize{}(6.21, 6.76)}\tabularnewline
 & {\footnotesize{}ESS/A (min, max)} & {\footnotesize{}(0.31, 0.34) } &  & {\footnotesize{}(1.09, 1.58) } & {\footnotesize{}(1.34, 2.80) } &  & {\footnotesize{}(1.60, 1.60) } & {\footnotesize{}(2.53, 2.76) }\tabularnewline
\hline 
\end{tabular}{\footnotesize\par}
\par\end{centering}
\caption{\label{tab:Effective-sample-sizes-wishart}Effective sample sizes
and computing times for the dynamic inverted Wishart model (\ref{eq:wishart_1},\ref{eq:wishart_4}).
In all cases, the results are based on 10 independent chains/trajectories
and reported computing times are the total computing times over these
10 replica. For rstan, default sampler parameters with 1000 warmup
transitions followed by 1000 sampling transitions were used. For pdphmc,
$T=5000$ split evenly between warmup and sampling, 1000 samples and
an VARI type diagonal mass matrix were applied. Both computing times
for the sampling (S) period and for all (A) computations (warmup and
sampling) are provided, and ESSes are weighted both for S and A.}
\end{table}
As a large scale illustrative application of NGRHMC, the dynamic inverted
Wishart model for realized covariance matrices \citep{Golosnoy2012211}
of \citet{1601.01125} is considered. Under this model, a time series
of SPD covariance matrices $\mathbf{Y}_{k}\in\mathbb{R}^{G\times G},\;k=1,\dots,n$
are modeled independently inverted Wishart distributed conditionally
on a latent time-varying SPD scale matrix $\boldsymbol{\Sigma}_{k}$
and a degree of freedom parameter $\nu>G+1$, i.e. 
\begin{equation}
\mathbf{Y}_{k}|\boldsymbol{\Sigma}_{k},\nu\sim\text{inv-Wishart}(\nu,\boldsymbol{\Sigma}_{k})\label{eq:wishart_1}
\end{equation}
so that $E(\mathbf{Y}_{k}|\boldsymbol{\Sigma}_{k},\nu)=(\nu-G-1)^{-1}\boldsymbol{\Sigma}_{k}$.
The time-varying scale matrix is in turn specified in terms of 
\begin{equation}
\boldsymbol{\Sigma}_{k}=\mathbf{H}[\text{diag}(\exp(\mathbf{x}_{1,k}),\dots,\exp(\mathbf{x}_{G,k}))]\mathbf{H}^{T}\label{eq:wishart_2}
\end{equation}
where $\mathbf{H}\in\mathbb{R}^{G\times G}$ is a lower triangular
matrix with $\mathbf{H}_{g,g}=1,\;g=1,\dots,G$. The remaining (strictly
lower triangular) elements $\mathbf{H}_{i,j},\;j=1,\dots,G-1,i=j+1,\dots,G,$
are unrestricted parameters. Finally, the log-scale factors $\mathbf{x}_{g,k}$
are a-priori independent (over $g$) stationary Gaussian AR(1) processes
\begin{align}
\mathbf{x}_{g,k} & =\boldsymbol{\mu}_{g}+\boldsymbol{\delta}_{g}(\mathbf{x}_{g,k-1}-\boldsymbol{\mu}_{g})+\boldsymbol{\sigma}_{g}\varepsilon_{g,k},\;\varepsilon_{g,k}\sim\text{iid }N(0,1),\;k=2,\dots,n,\;g=1,\dots,G,\label{eq:wishart_3}\\
\mathbf{x}_{g,1} & \sim N\left(\boldsymbol{\mu}_{g},\boldsymbol{\sigma}_{g}^{2}/(1-\boldsymbol{\delta}_{g}^{2})\right),\;g=1,\dots,G,\label{eq:wishart_4}
\end{align}
where \textbf{$\boldsymbol{\mu}_{g},\;\boldsymbol{\delta}_{g}\in(-1,1),\;\boldsymbol{\sigma}_{g}>0,\;g=1,\dots,G$
}are parameters. 

The joint distribution of parameters $\boldsymbol{\theta}=(\boldsymbol{\mu},\boldsymbol{\delta},\boldsymbol{\sigma},\mathbf{H}_{2:G,1},\dots,\mathbf{H}_{G,G-1},\nu)$
and latent variables $\mathbf{x}$ is difficult to sample from, and
in order to reduce ``funnel'' effects, the Laplace-based transport
map reformulation of \citet{1812.07929} (without Newton iterations)
is used here. For every admissible $\boldsymbol{\theta}$, a smooth
bijective mapping, say $\mathbf{\mathbf{x}}=\text{\ensuremath{\gamma}}_{\boldsymbol{\theta}}(\mathbf{z}),\;\mathbf{z}\in\mathbb{R}^{Gn}$,
is introduced so that $p(\mathbf{z},\boldsymbol{\theta}|\mathbf{Y}_{1:n})\propto|\nabla_{\mathbf{z}}\text{vec}(\text{\ensuremath{\gamma}}_{\boldsymbol{\theta}}(\mathbf{z}))|[p(\mathbf{x},\boldsymbol{\theta}|\mathbf{Y}_{1:n})]_{\mathbf{\mathbf{x}}=\text{\ensuremath{\gamma}}_{\boldsymbol{\theta}}(\mathbf{z})}$
approximates $p(\boldsymbol{\theta}|\mathbf{Y}_{1:n})\mathcal{N}(\mathbf{z}|\mathbf{0}_{Gn},\mathbf{I}_{Gn})$.
Subsequently, NGRHMC/HMC targeting $p(\mathbf{z},\boldsymbol{\theta}|\mathbf{Y}_{1:n})$
are performed. The reader is referred Appendix \ref{subsec:details__dynamic-inverted-Wishart}
and \citet{1812.07929} for more details on the construction of $\text{\ensuremath{\gamma}}_{\boldsymbol{\theta}}$
and Appendix \ref{subsec:details__dynamic-inverted-Wishart} for further
details such as priors.

The data considered are $n=2514$ daily observations of realized covariance
matrices for $G=5$ stocks (American Express, Citigroup, General Electric,
Home Depot, IBM) between January 1, 2000 and December 31, 2009. See
\citet{Golosnoy2012211} for details on how this dataset was constructed
from high frequency financial data. For these values of $n$ and $G$,
the model involves $Gn=12570$ latent variables and $3G+G(G-1)/2+1=26$
parameters.

\subsection{Results}

ESSes and time-weighted ESSes for the parameters, $\mathbf{z}_{1:5,1}$
and $\mathbf{x}_{1:5,1}$ are given in Table \ref{tab:Effective-sample-sizes-wishart}
for two variants of pdphmc and an rstan benchmark. It is seen that
the discretely sampled (D) pdphmc uniformly provides faster sampling
performance than rstan. The speedup is in particularly significant
when taking all (A) computations into account as rstan uses more than
80\% of the computing time in the warmup phase, whereas there is relatively
little warmup overhead for pdphmc. For continuously sampled (C) pdphmc
the picture is somewhat more mixed with numbers ranging from being
on par with rstan ($\boldsymbol{\delta}$, event specification 2)
to being up to five times faster ($\mathbf{H}$, both event specifications).
Event specification 2 lead to better worst case performance than event
specification 1 in all cases other than for $\mathbf{H}$. However,
the differences are not very large which may be explained by the high
dimensionality of the model, and that event specifications 1 and 2
are very similar in this case. Still this observation suggest that
it may be possible to gain even more efficiency by developing better
adaptive event rate specifications. 

Posterior means and standard deviations obtained both from rstan and
discretely sampled pdphmc presented in Table \ref{tab:Posterior-means-and-wishart}
in the Appendix show no noteworthy deviations \citep[see also][Table 5]{1601.01125}.
To conclude, pdphmc is fast and reliable alternative to HMC that also
scale well to high-dimensional settings. 

\section{Discussion\label{sec:Discussion}}

This paper has introduced Numerical Generalized Randomized HMC processes
as a new, robust and potentially very efficient alternative to conventional
MCMC methods. The presently proposed methodology holds promise to
be substantially more trustworthy for complicated real-life problems.
This improvement is related to two factors:
\begin{itemize}
\item The NGRHMC process is time-irreversible, and the present paper is
to the author's knowledge the among the first attempts to leverage
time-irreversible processes to produce general purpose and easy to
use MCMC-like samplers that scale to high-dimensional problems. By
now, there is substantial evidence \citep[see e.g. discussion on page 387 of][]{fearnhead2018}
that irreversible processes are superior to conventional reversible
alternatives such as HMC. In particular, irreversible methods appear
to be more robust to irregular target distributions, as transition
can be made without regard to the likelihood of the corresponding
reversed transition occurring.
\item The proposed implementation of NGRHMC process leverages the mature,
and widely used field of numerical integration of ordinary differential
equations. Common practice for HMC is choosing a fixed step size low
order symplectic method and hoping that regions where this step size
is too large for numerical stability is not encountered during the
simulation. The proposed methodology, on the other hand, relies on
high quality adaptive integrators, which have no such stability problems. 
\end{itemize}
Currently, efficient and robust MCMC computations has been a field
dominated by tailor-making to specific applications and a large degree
of craftsmanship. Effectively, the two above points reduces such MCMC
computations into a more routine task of numerically integrating ordinary
differential equations using adaptive/automatic methods.

There scope for substantial further work on NGRHMC-processes beyond
the initial developments given here. A, by no means complete, list
of possible further research directions related to NGRHMC processes
is given in Appendix \ref{sec:Further-work}.

\bibliographystyle{chicago}
\bibliography{../../bibtex/kleppe}

\newpage{}

\appendix
\setcounter{page}{1}
\begin{center}
{\Large{}Online Appendix to ``Connecting the Dots: Numerical Randomized
Hamiltonian Monte Carlo with State-Dependent Event Rates'' by Tore
Selland Kleppe} 
\par\end{center}

This version: \today\\

\begin{flushleft}
In this online Appendix, equation numbers $\leq$ \ref{eq:wishart_4}
refer to equations in the main article text. 
\par\end{flushleft}

\section{Details of derivations \label{sec:Details-of-derivations}}

\subsection{The left-hand side of (\ref{eq:fokker_planck}) is the Poisson bracket
between $p$ and $\mathcal{H}$\label{subsec:Poisson_bracket}}

Suppose $D=2d$, $\mathbf{z}=[\mathbf{q}^{T},\mathbf{p}^{T}]^{T}$
and $\Phi(\mathbf{z})$ corresponds to Hamilton's equations associated
with separable Hamiltonian (\ref{eq:Hamiltonian}), i.e.
\begin{align*}
\Phi(\mathbf{z}) & =\left[\begin{array}{c}
\mathbf{M}^{-1}\mathbf{p}\\
\nabla_{\mathbf{q}}\log\tilde{\pi}(\mathbf{q})
\end{array}\right]=\mathbf{J}\nabla_{\mathbf{z}}\mathcal{H}(\mathbf{z}),\;\mathbf{J}=\left[\begin{array}{cc}
\mathbf{0}_{d,d} & \mathbf{I}_{d}\\
-\mathbf{I}_{d} & \mathbf{0}_{d,d}
\end{array}\right],
\end{align*}
Then the left-hand side of the stationary Fokker-Planck equation (\ref{eq:fokker_planck})
is equal to the Poisson Bracket $\left\{ p,\mathcal{H}\right\} (\mathbf{z})$
\citep[see e.g.][Section 3.3]{Leimkuhler:2004} between the density
$p(\mathbf{z})$ and the Hamiltonian $\mathcal{H}(\mathbf{z})$, namely
\begin{align*}
\left\{ p,\mathcal{H}\right\} (\mathbf{z}) & :=\left[\nabla_{\mathbf{z}}p(\mathbf{z})\right]^{T}\mathbf{J}\nabla_{\mathbf{z}}\mathcal{H}(\mathbf{z}),\\
 & =\left[\nabla_{\mathbf{q}}p(\mathbf{z})\right]^{T}\mathbf{M}^{-1}\mathbf{p}+\left[\nabla_{\mathbf{p}}p(\mathbf{z})\right]^{T}\nabla_{\mathbf{q}}\log\tilde{\pi}(\mathbf{q}),\\
 & =\sum_{i=1}^{d}\frac{\partial}{\partial q_{i}}[\mathbf{M}^{-1}\mathbf{p}]_{i}p(\mathbf{z})+\sum_{i=1}^{d}\frac{\partial}{\partial p_{i}}[\nabla_{\mathbf{q}}\log\tilde{\pi}(\mathbf{q})]_{i}p(\mathbf{z}),\\
 & =\sum_{i=1}^{D}\frac{\partial}{\partial z_{i}}\left[\Phi_{i}(\mathbf{z})p(\mathbf{z})\right].
\end{align*}
For any first integral, say $g(\mathbf{z})$ (i.e. conserved quantity
so that $g(\varphi_{t}(\mathbf{z}))=g(\mathbf{z})\;\forall\;t,\mathbf{z}$),
the Poisson bracket between $g$ and $\mathcal{H}$ is zero for all
$\mathbf{z}$. Clearly, the BG distribution $\rho(\mathbf{z})\propto\exp(-\mathcal{H}(\mathbf{z}))$
is a first integral of $\mathcal{H}$, and hence $\left\{ \mathcal{H},\rho\right\} (\mathbf{z})=\sum_{i=1}^{D}\frac{\partial}{\partial z_{i}}\left[\Phi_{i}(\mathbf{z})\rho(\mathbf{z})\right]=0\;\forall\;\mathbf{z}$. 

Note also that the same argument goes through for a general separable
Hamiltonian, say $\mathcal{H}(\mathbf{z})=-\nabla_{\mathbf{q}}\log\tilde{\pi}(\mathbf{q})+V(\mathbf{p})$,
with $\Phi(\mathbf{z})=[\nabla_{\mathbf{p}}V(\mathbf{p})^{T},\nabla_{\mathbf{q}}\log\tilde{\pi}(\mathbf{q})]$.

\subsection{General Markov kernel momentum refreshes\label{subsec:General-Markov-kernel}}

Given some event rate $\lambda(\mathbf{q},\mathbf{p})>0$ so that
$C(\mathbf{q})=\int\mathcal{N}(\mathbf{p}|\mathbf{0}_{d},\mathbf{M})\lambda(\mathbf{q},\mathbf{p})d\mathbf{p}<\infty$
for each admissible $\mathbf{q}$. Define the conditional density
\[
v_{\mathbf{q}}(\mathbf{p})=\frac{\mathcal{N}(\mathbf{p}|\mathbf{0}_{d},\mathbf{M})\lambda(\mathbf{q},\mathbf{p})}{C(\mathbf{q})},
\]
and, for each fixed $\mathbf{q}$, let $K_{\mathbf{q}}(\mathbf{p}|\mathbf{p}^{\prime})$
be the density of a Markov kernel that leaves $v_{\mathbf{q}}(\mathbf{p})$
invariant, i.e.
\[
\int K_{\mathbf{q}}(\mathbf{p}|\mathbf{p}^{\prime})v_{\mathbf{q}}(\mathbf{p}^{\prime})d\mathbf{p}^{\prime}=v_{\mathbf{q}}(\mathbf{p}).
\]
Then, take the transition distribution at events to be 
\begin{equation}
Q(\mathbf{z}|\mathbf{z}^{\prime})=\delta(\mathbf{q}-\mathbf{q}^{\prime})K_{\mathbf{q}^{\prime}}(\mathbf{p}|\mathbf{p}^{\prime}).\label{eq:general_Q-1}
\end{equation}
Under this event specification, the right-hand side of (\ref{eq:fokker_planck})
(with $p(\mathbf{z})=\rho(\mathbf{z})$) reduces to 
\begin{align*}
 & \int\pi(\mathbf{q}^{\prime})\mathcal{N}(\mathbf{p}^{\prime}|\mathbf{0}_{d},\mathbf{M})\lambda(\mathbf{q}^{\prime},\mathbf{p}^{\prime})\delta(\mathbf{q}-\mathbf{q}^{\prime})K_{\mathbf{q^{\prime}}}(\mathbf{p}|\mathbf{p}^{\prime})d\mathbf{z}^{\prime}-\rho(\mathbf{z})\lambda(\mathbf{z}),\\
= & \pi(\mathbf{q})\int\mathcal{N}(\mathbf{p}^{\prime}|\mathbf{0}_{d},\mathbf{M})\lambda(\mathbf{q},\mathbf{p}^{\prime})K_{\mathbf{q}}(\mathbf{p}|\mathbf{p}^{\prime})d\mathbf{p}^{\prime}-\rho(\mathbf{z})\lambda(\mathbf{z}),\\
= & \pi(\mathbf{q})C(\mathbf{q})\int K_{\mathbf{q}}(\mathbf{p}|\mathbf{p}^{\prime})v_{\mathbf{q}}(\mathbf{p}^{\prime})d\mathbf{p}^{\prime}-\rho(\mathbf{z})\lambda(\mathbf{z}),\\
= & \pi(\mathbf{q})C(\mathbf{q})v_{\mathbf{q}}(\mathbf{p})-\rho(\mathbf{z})\lambda(\mathbf{z}),\\
= & \pi(\mathbf{q})\mathcal{N}(\mathbf{p}|\mathbf{0}_{d},\mathbf{M})\lambda(\mathbf{q},\mathbf{p})-\rho(\mathbf{z})\lambda(\mathbf{z})=0.
\end{align*}
I.e. provided the deterministic dynamics is Hamiltonian as described,
the PDMP the process with general event rate $\lambda(\mathbf{q},\mathbf{p})$
and transition distribution at events (\ref{eq:general_Q-1}) admit
$\rho(\mathbf{z})$ as a stationary distribution. Note also that the
above argument may be modified to accommodate a general $\mathbf{p}$-marginal
of the BG distribution. 

\subsection{Riemann manifold variants\label{sec:Riemann-manifold-variants}}

This section gives the relevant details for constructing continuous
time Riemann manifold HMC processes. Such processes rely on selecting
a symmetric positive definite ``metric tensor'' $\mathbf{G}(\mathbf{q})\in\mathbb{R}^{d\times d}$
which should reflect the ``local'' precision of the target distribution
around $\mathbf{q}$. The non-separable Hamiltonian $\mathcal{K}$
typically used in such situations is given by \citep{girolami_calderhead_11}
\[
\mathcal{K}(\mathbf{z})=-\log\tilde{\pi}(\mathbf{q})+\frac{1}{2}\log(|\mathbf{G}(\mathbf{q})|)+\frac{1}{2}\mathbf{p}^{T}\mathbf{G}(\mathbf{q})^{-1}\mathbf{p}.
\]
Hamilton's equations are still given by 
\[
\Phi(\mathbf{z})=\mathbf{J}\nabla_{\mathbf{z}}\mathcal{K}(\mathbf{z})=\left[\begin{array}{c}
\nabla_{\mathbf{p}}\mathcal{K}(\mathbf{z})\\
-\nabla_{\mathbf{q}}\mathcal{K}(\mathbf{z})
\end{array}\right],
\]
and BG distribution by $\rho(\mathbf{z})=\pi(\mathbf{q})\mathcal{N}(\mathbf{p}|\mathbf{0},\mathbf{G}(\mathbf{q}))=\exp(-\mathcal{K}(\mathbf{z}))W^{-1}$,
$W=\int\exp(-\mathcal{K}(\mathbf{z}))d\mathbf{z}$. In terms of numerical
implementation, it is seen that Hamilton's equations cannot in this
case be reduced to a second order ODE but may still be solved using
general purpose first order ODE solvers such as Runge-Kutta methods.
A drawback of general Riemann manifold HMC methods is that symplectic
integrators for non-separable Hamiltonians are necessary implicit,
and hence require many evaluations of $\Phi$ per integration step.
A pro of the proposed methodology is that adaptive explicit solvers
may be used.

The strategy for showing that the drift term in the Fokker-Planck
equation (\ref{eq:fokker_planck}) vanishes also with Hamiltonian
$\mathcal{K}$ is to look directly at the drift term with $p=\rho$.
In this case
\begin{align*}
\sum_{i=1}^{D}\frac{\partial}{\partial z_{i}}\left[\Phi_{i}\rho(\mathbf{z})\right]= & W^{-1}\sum_{i=1}^{d}\frac{\partial}{\partial q_{i}}\left[\exp(-\mathcal{K}(\mathbf{z}))\frac{\partial}{\partial p_{i}}\mathcal{K}(\mathbf{z})\right]\\
 & -W^{-1}\sum_{i=1}^{d}\frac{\partial}{\partial p_{i}}\left[\exp(-\mathcal{K}(\mathbf{z}))\frac{\partial}{\partial q_{i}}\mathcal{K}(\mathbf{z})\right],\\
= & -\rho(\mathbf{z})\sum_{i=1}^{d}\left[\frac{\partial}{\partial q_{i}}\mathcal{K}(\mathbf{z})\right]\left[\frac{\partial}{\partial p_{i}}\mathcal{K}(\mathbf{z})\right]\\
 & +\rho(\mathbf{z})\sum_{i=1}^{d}\frac{\partial^{2}}{\partial q_{i}\partial p_{i}}\mathcal{K}(\mathbf{z})\\
 & +\rho(\mathbf{z})\sum_{i=1}^{d}\left[\frac{\partial}{\partial p_{i}}\mathcal{K}(\mathbf{z})\right]\left[\frac{\partial}{\partial q_{i}}\mathcal{K}(\mathbf{z})\right]\\
 & -\rho(\mathbf{z})\sum_{i=1}^{d}\frac{\partial^{2}}{\partial p_{i}\partial q_{i}}\mathcal{K}(\mathbf{z}),\\
= & 0.
\end{align*}
(For clarity; the second equality is based on the chain rule $\frac{\partial}{\partial z_{i}}\exp(-\mathcal{K}(\mathbf{z}))\Phi_{i}(\mathbf{z})=-\exp(-\mathcal{K}(\mathbf{z}))\Phi_{i}(\mathbf{z})\frac{\partial}{\partial z_{i}}\mathcal{K}(\mathbf{z})$
$+\exp(-\mathcal{K}(\mathbf{z}))\frac{\partial}{\partial z_{i}}\Phi_{i}(\mathbf{z)}$). 

As also the flow associated with $\mathcal{K}$ preserves the BG distribution,
one is also in this case free to choose the event specification solely
on making the right-hand side of (\ref{eq:fokker_planck}) vanish.
Essentially the same event specification cases as in the separable
case, with minimal modifications to account for the position-dependence
in the BG distribution $\mathbf{p}$-marginal, applies. I.e. 
\[
\lambda(\mathbf{z})=\omega(\mathbf{q}),\;\;Q(\mathbf{p}|\mathbf{z}^{\prime})=\mathcal{N}(\mathbf{p}|\phi\mathbf{p}^{\prime},\sqrt{1-\phi^{2}}\mathbf{G}(\mathbf{q}^{\prime})),
\]
and
\[
\lambda(\mathbf{z})=\lambda(\mathbf{q},\mathbf{p}),\;\;Q(\mathbf{p}|\mathbf{z}^{\prime})\propto\lambda(\mathbf{q}^{\prime},\mathbf{p})\mathcal{N}(\mathbf{p}|\mathbf{0}_{d},\mathbf{G}(\mathbf{q}^{\prime})),
\]
lead to on target processes. The arguments leading to this conclusion
are analogous to those in Appendix \ref{subsec:General-Markov-kernel}
and are not repeated.

\subsection{Langevin limit\label{sec:Langevin-limit}}

To obtain the Langevin limit under event specification 1, consider
first a large, but finite $\lambda$, and condition on a sequence
of inter-event times $\tau_{i}\sim\text{ iid }Exp(\lambda^{-1})\;i=1,2,\dots$.
A time-discretized version of the PDMP $\mathcal{Z}_{t}$, say $\tilde{\mathbf{z}}_{j}=[\tilde{\mathbf{q}}_{j}^{T},\tilde{\mathbf{p}}_{j}^{T}]^{T}\approx\mathcal{Z}_{\sum_{i=1}^{j}\tau_{i}}$,
obtains by first applying a single leap-frog step (with time step
$\tau_{i}$) to each deterministic transition to get the state immediately
before the $(i+1$)th event, say $\tilde{\mathbf{z}}_{i+1}^{*}$:
\begin{align}
\tilde{\mathbf{p}}_{i+1/2} & =\tilde{\mathbf{p}}_{i}+\frac{\tau_{i+1}}{2}\nabla_{\mathbf{q}}\log\tilde{\pi}(\tilde{\mathbf{q}}_{i}),\label{eq:langevin_leap_1}\\
\tilde{\mathbf{q}}_{i+1}^{*} & =\tilde{\mathbf{q}}_{i}+\tau_{i+1}\mathbf{M}^{-1}\tilde{\mathbf{p}}_{i+1/2},\label{eq:langevin_leap_2}\\
\tilde{\mathbf{p}}_{i+1}^{*} & =\tilde{\mathbf{p}}_{i+1/2}+\frac{\tau_{i+1}}{2}\nabla_{\mathbf{q}}\log\tilde{\pi}(\tilde{\mathbf{q}}_{i+1}^{*}).\nonumber 
\end{align}
(the latter equation is not needed but is given for completeness of
the leapfrog step). Secondly, at the event (at time $\sum_{j=1}^{i+1}\tau_{j}$)
$\tilde{\mathbf{z}}_{i+1}$ is sampled according to $Q(\cdot|\tilde{\mathbf{z}}_{i+1}^{*})$,
which reduces to:
\begin{align}
\tilde{\mathbf{q}}_{i+1} & =\tilde{\mathbf{q}}_{i+1}^{*}\label{eq:langevin_sample_1}\\
\tilde{\mathbf{p}}_{i+1} & =\varepsilon_{i+1},\;\varepsilon_{i+1}\sim\text{ iid }N(\mathbf{0}_{d,1},\mathbf{M}).\label{eq:langevin_sample_2}
\end{align}
Now, since the time-discrete momentum is resampled each step, the
discrete time (marginal) position dynamics $\tilde{\mathbf{q}}_{i}$
obtains by combining (\ref{eq:langevin_leap_1},\ref{eq:langevin_sample_1})
and time-shifted (\ref{eq:langevin_sample_2}) into (\ref{eq:langevin_leap_2}):
\begin{equation}
\tilde{\mathbf{q}}_{i+1}=\tilde{\mathbf{q}}_{i}+\frac{\tau_{i+1}^{2}}{2}\mathbf{M}^{-1}\nabla_{\mathbf{q}}\log\tilde{\pi}(\tilde{\mathbf{q}}_{i})+\tau_{i}\mathbf{M}^{-\frac{1}{2}}\eta_{i},\label{eq:Langevin-euler}
\end{equation}
where $\eta_{i}=\mathbf{M}^{-\frac{1}{2}}\varepsilon_{i}\sim N(\mathbf{0}_{d,1},\mathbf{I}_{d})$.
Equation \ref{eq:Langevin-euler} is recognized to be the Euler-Maruyama
discretization of (\ref{eq:langevin_limit}) with time step size $\tau_{i+1}^{2}$.
Finally, upon letting $\lambda\rightarrow\infty$, the discrete time
process $\tilde{\mathbf{z}}_{j}$ converges to the underlying PDMP,
and $\tilde{\mathbf{q}}_{j}$ to the Langevin process (\ref{eq:langevin_limit})
a.s.

\subsection{Arc-length and event specification 2\label{subsec:Arc-length-and-event}}

Consider the calculation of the arc-length of some (position) dynamics
$\mathbf{q}(s),\;s\in[0,t]$ using the distance
\[
d(\mathbf{q},\mathbf{q}^{\prime})=\sqrt{(\mathbf{q}-\mathbf{q}^{\prime})^{T}\mathbf{M}(\mathbf{q}-\mathbf{q}^{\prime})}.
\]
Notice that throughout this text, $\mathbf{M}^{-1}$ should reflect
$Var(\mathbf{q})$ and hence the distance $d$ aims at making the
contribution of each dimension more uniform. Discretizing time $s_{i}=i\Delta$,
$\Delta=t/N$, the arc-length may be arbitrarily well approximated
by
\begin{align*}
A_{N} & =\sum_{i=1}^{N}\sqrt{(\mathbf{q}(s_{i})-\mathbf{q}(s_{i-1}))^{T}\mathbf{M}(\mathbf{q}(s_{i})-\mathbf{q}(s_{i-1}))},\\
 & =\sum_{i=1}^{N}\Delta\sqrt{\left[\frac{\mathbf{q}(s_{i})-\mathbf{q}(s_{i-1})}{\Delta}\right]^{T}\mathbf{M}\left[\frac{\mathbf{q}(s_{i})-\mathbf{q}(s_{i-1})}{\Delta}\right]},
\end{align*}
for large $N$. Letting $N\rightarrow\infty$ one obtains by a standard
limit argument that 
\[
A_{N}\rightarrow A=\int_{0}^{t}\sqrt{\left[\dot{\mathbf{q}}(s)\right]^{T}\mathbf{M}\dot{\mathbf{q}}(s)}ds.
\]
Finally, from the first of Hamilton's equations, $\dot{\mathbf{q}}(s)=\mathbf{M}^{-1}\mathbf{p}(s)$:
\[
A=\int\sqrt{\mathbf{p}(s)^{T}\mathbf{M}^{-1}\mathbf{p}(s)}ds
\]
which is (up to a multiplicative constant) the integrated $\lambda(\mathbf{z})$
(\ref{eq:integrated_event_intensity}) used in event specification
2.

\section{Temporal averages of the Hamiltonian dynamics for Gaussian targets\label{sec:Temporal-averages-of}}

This section considers the temporal average of the flow for \emph{Gaussian}
targets. In particular it is shown that exploring only a single energy
levelset is sufficient to get asymptotically correct estimates for
any linear combination of the mean of the target when continuous sampling
is employed. 

Consider a $N(\mu,\Sigma)$ target distribution, where $\Sigma$ is
positive definite and finite. The Hamiltonian dynamics $\mathbf{z}(t)$
is then the solution to the linear differential equation 
\begin{equation}
\dot{\mathbf{z}}(t)=\mathbf{B}\mathbf{z}(t)+\mathbf{b}\label{eq:gauss_lin_ode}
\end{equation}
where 
\[
\mathbf{B}=\left[\begin{array}{cc}
\mathbf{0}_{d,d} & \mathbf{M}^{-1}\\
-\Sigma^{-1} & \mathbf{0}_{d,d}
\end{array}\right],\;\mathbf{b}=\left[\begin{array}{c}
\mathbf{0}_{d,1}\\
\Sigma^{-1}\mu
\end{array}\right].
\]
Consider first the case where $\mu=\mathbf{0}_{d}$, which results
in the solution to (\ref{eq:gauss_lin_ode}) given by $\mathbf{z}(t)=\exp(t\mathbf{B})\mathbf{z}(0)$
where $\exp(t\mathbf{B})$ is the matrix exponential of matrix $t\mathbf{B}$.
This solution necessarily conserves the associated Hamiltonian, which
may be written as
\begin{equation}
\mathcal{H}(\mathbf{z}(t))=\frac{1}{2}\mathbf{z}(t)^{T}\mathbf{B}^{*}\mathbf{z}(t),\;\mathbf{B}^{*}=\left[\begin{array}{cc}
\Sigma^{-1} & \mathbf{0}_{d,d}\\
\mathbf{0}_{d,d} & \mathbf{M}^{-1}
\end{array}\right].\label{eq:hamiltonian_elliptical}
\end{equation}
From the conservation of (\ref{eq:hamiltonian_elliptical}), it is
clear that the solutions $\mathbf{z}(t)$ are restricted to an ellipsoid
(centered in $\mathbf{0}_{2d,1}$) in $\mathbb{R}^{2d}$, and hence
$\mathbf{\exp}(t\mathbf{B})\mathbf{v}=O(1)$ as $t\rightarrow\infty$
for any finite vector $\mathbf{v}\in\mathbb{R}^{2d}$. 

Now, consider the general $\mu\in\mathbb{R}^{d}$ case. Any linear
algebra textbook provides the solution to (\ref{eq:gauss_lin_ode}),
in terms of an initial state $\mathbf{z}(0)$, namely
\[
\mathbf{z}(t)=\exp(t\mathbf{B})\mathbf{z}(0)+\mathbf{B}^{-1}\left[\exp(t\mathbf{B})-\mathbf{I}_{2d}\right]\mathbf{b}.
\]
Suppose now that we seek the temporal average of some linear combination,
say $\mathbf{K}\mathbf{z}(t)$, $\mathbf{K}\in\mathbb{R}^{l\times2d}$,
of the dynamics:
\begin{align*}
\frac{1}{T}\int_{0}^{T}\mathbf{K}\mathbf{z}(t)dt= & \frac{1}{T}\mathbf{K}\left[\int_{0}^{T}\exp(t\mathbf{B})dt\right]\mathbf{z}(0)+\frac{1}{T}\mathbf{K}\mathbf{B}^{-1}\left[\int_{0}^{T}\exp(t\mathbf{B})dt\right]\mathbf{b}-\mathbf{K}\mathbf{B}^{-1}\mathbf{b}\\
= & \frac{1}{T}\mathbf{K}\mathbf{B}^{-1}[\exp(T\mathbf{B})-\mathbf{I}_{2d}]\mathbf{z}(0)\\
 & +\frac{1}{T}\mathbf{K}\mathbf{B}^{-1}\mathbf{B}^{-1}[\exp(T\mathbf{B})-\mathbf{I}_{2d}]\mathbf{b}\\
 & +\mathbf{K}[\mu^{T},\mathbf{0}_{d,1}^{T}]^{T}.
\end{align*}
Now, former two terms in the latter representation of $\frac{1}{T}\int_{0}^{T}\mathbf{K}\mathbf{z}(t)dt$
vanishes as $T\rightarrow\infty$, (since $\mathbf{K}\mathbf{B}^{-1}[\exp(T\mathbf{B})-\mathbf{I}_{2d}]\mathbf{z}(0)=O(1)$
and $\mathbf{K}\mathbf{B}^{-1}\mathbf{B}^{-1}[\exp(T\mathbf{B})-\mathbf{I}_{2d}]\mathbf{b}=O(1)$),
and one may conclude that
\[
\frac{1}{T}\int_{0}^{T}\mathbf{K}\mathbf{z}(t)dt\underset{T\rightarrow\infty}{\longrightarrow}\mathbf{K}\left[\begin{array}{c}
\mu\\
\mathbf{0}_{d,1}
\end{array}\right],
\]
invariantly of $\mathbf{z}(0)$ (or equivalently the energy level
set $\mathbf{z}(t)$ is restricted to). 

For the interested reader; a further special case obtains when the
mass matrix is chosen to be the precision of the target, i.e. $\mathbf{M}=\Sigma^{-1}$.
Then the centered flow takes a particularly simple form:
\[
\exp(t\mathbf{B})=\left[\begin{array}{cc}
\cos(t)\mathbf{I}_{d,d} & \sin(t)\Sigma\\
-\sin(t)\Sigma & \cos(t)\mathbf{I}_{d,d}
\end{array}\right].
\]
(This latter expression obtains from that even powers of $\mathbf{B}$
are $\pm\mathbf{I}_{2d}$ and odd powers of $\mathbf{B}$ are $\pm\mathbf{B}$,
and subsequently employing this information in infinite series defining
$\exp(t\mathbf{B})$.) In this case, the former two terms in the latter
representation of $\frac{1}{T}\int_{0}^{T}\mathbf{K}\mathbf{z}(t)dt$
above vanishes whenever $t=2\pi k$, $k$ is an integer. I.e. the
trajectory forms exactly $k$ closed orbits centered in $\mu$ on
the underlying ellipsoid, and the average then becomes $\mu$. For
$\mathbf{M}\neq\Sigma^{-1}$, the time $T_{i}$ required for $T_{i}^{-1}\int_{0}^{T_{i}}q_{i}(t)dt=\mu_{i}$
generally depends on $i$, and thus only the asymptotic result above
holds in this case.

\begin{comment}

\subsection{Quadratic forms in Gaussian flows}

Of interest in the Monte Carlo studies involving a Gaussian target
distribution is the ability to calculated quadratic forms of the type
(restricting to the $\mu=\mathbf{0}_{d}$ case)
\[
\int_{0}^{T}\mathbf{z}(t)^{T}\mathbf{U}\mathbf{z}(t)dt
\]
where $\mathbf{U}$ is a symmetric positive semidefinite matrix, and
$\mathbf{z}(t)=\exp(t\mathbf{B})\mathbf{z}(0)$ as above. The integral
may be rewritten \citep{vanloan_exp_int} as
\[
\mathbf{z}(0)^{T}\int_{0}^{T}\exp(t\mathbf{B}^{T})\mathbf{U}\exp(t\mathbf{B})dt\mathbf{z}(0)=\mathbf{z}(0)^{T}\mathbf{A}_{3}(T)^{T}\mathbf{A}_{2}(T)\mathbf{z}(0)
\]
where 
\[
\exp\left(T\left[\begin{array}{cc}
-\mathbf{B}^{T} & \mathbf{U}\\
\mathbf{0}_{2d,2d} & \mathbf{B}
\end{array}\right]\right)=\left[\begin{array}{cc}
\mathbf{A}_{1}(T) & \mathbf{A}_{2}(T)\\
\mathbf{0}_{2d,2d} & \mathbf{A}_{3}(T)
\end{array}\right].
\]

\end{comment}

\section{Details related to numerical implementation\label{sec:Details-related-to-implementation}}

Denote by $\psi_{\varepsilon}$ the RKN step. Then $\hat{\mathbf{s}}(\tau+\varepsilon)=\psi_{\varepsilon}(\mathbf{s}(\tau))$,
starting at $\mathbf{s}(\tau)$ and with time step size $\varepsilon$,
is a 6th order approximation of the flow. In addition, when $\psi_{\varepsilon}$
has been evaluated, the following is available at negligible additional
cost (in particular no further evaluations of the right-hand side
of (\ref{eq:augmented_sec_order_ode})):
\begin{enumerate}
\item A 4th order approximation, say $\tilde{\mathbf{s}}(\tau+\varepsilon)$,
of $\mathbf{s}(\tau+\varepsilon)$ used for estimation of the (local)
error incurred, which is subsequently used in the adaptive step size
selection.
\item A 6th order interpolation approximation, say $\hat{\mathbf{s}}(\tau+\xi\varepsilon)$,
of any element of $(\mathbf{q}(\tau+\xi\varepsilon),\dot{\mathbf{q}}(\tau+\xi\varepsilon),\mathbf{r}(\tau+\xi\varepsilon)),\;\xi\in(0,1)$. 
\end{enumerate}
A fairly standard PI-step size controller \citep[see e.g.][Section 17.2.1]{numrecipes2007}
was used to select $\varepsilon$ dynamically, by maintaining that
\[
\max_{i}\frac{|\hat{\mathbf{s}}_{i}(\tau+\varepsilon)-\tilde{\mathbf{s}}_{i}(\tau+\varepsilon)|}{tol_{a}+tol_{r}|\hat{\mathbf{s}}_{i}(\tau+\varepsilon)|}<1
\]
for all time RKN steps. Here, $tol_{a}$ and $tol_{r}$ are the absolute-
and relative error tolerances respectively, which must be chosen a-priori.

\begin{algorithm}
\begin{algorithmic}
\State {\bf Input:} Target log-density kernel $\log \tilde \pi(\mathbf q)$, $\mathbf q \in \mathbb R^d$ associated with target density $\pi(\mathbf q)$. 
\State {\bf Input:} Event specification $(\lambda,Q)$ (see e.g. Table \ref{tab:The-event-specifications}). 
\State {\bf Input:} Mass matrix $\mathbf M\in\mathbb R^{d \times d}$, symmetric, positive definite (e.g. $\mathbf M=\mathbf I_d$).
\State {\bf Input:} Simulation time span $T\in \mathbb R^+$.
\State {\bf Input:} Number of discrete time samples $N$.
\State {\bf Input:} Initial position $\mathbf q_0 \in \mathbb R^d$.
\State {\bf Input:} Numerical error tolerances $tol_a,\;tol_r$ (e.g. $tol_a=tol_r=1.0e-3$).
\State {\bf Input:} Moments $g_1,\dots,g_m$, $g_k:\mathbb R^d \mapsto \mathbb R,\;k=1,\dots,m$, for which $\int g_k(\mathbf q)\pi(\mathbf q)d\mathbf q$ are to be estimated. 
\State
\State Define integrated quantities $\mathscr{M}(\mathbf q) = [\lambda(\mathbf q),g_1(\mathbf q),\dots,g_m(\mathbf q)]:\mathbb R^d \mapsto \mathbb R ^{m+1}$.
\State $t\leftarrow 0$, 
\Comment{Continuous time process time.}
\State $\mathcal Q(0) \leftarrow \mathbf q_0$.
\Comment{Initial $\mathbf q$ configuration.}
\State $\mathcal P(0)\sim N(\mathbf 0_d,\mathbf M)$.
\Comment{Initial $\mathbf p$ configuration (in case of autocorrelated momentum refreshes)}
\State $\mathbf g \leftarrow \mathbf 0_m$ 
\Comment{Storage for estimated moments.}
\State $\mathcal S \leftarrow \mathbf 0_{d,M}$
\Comment{Storage for discrete time samples.}
\State $i \leftarrow 1$
\Comment{Discrete samples counter.}
\While{$t<T$}
\Comment{Main loop over events}
\State $\mathcal P(t) \sim Q(\cdot | \mathcal Z(t-))$.
\Comment{Simulate new momentum}
\State $u \sim Exp(1).$ 
\Comment{See equation \ref{eq:integrated_event_intensity}}
\State $\hat{\mathbf s}(0)\leftarrow (\mathcal Q(t),\mathbf M^{-1}\mathcal P(t),\mathbf 0_{m+1})$.
\Comment{Initial conditions for system of ODEs (\ref{eq:augmented_sec_order_ode})}
\State $\tau \leftarrow 0$. 
\Comment{Hamiltonian dynamics (between-events) time.}
\While{$\hat{\mathbf s}_{2d+1}(\tau) < u$ and $t+\tau<T$} \Comment{Recall $\hat{\mathbf s}_{2d+1}(\tau) \approx \int_0^\tau \lambda(\mathbf q(v))dv$}
\State Propose a new step size $\varepsilon$. 
\Comment{Using e.g. PI step size controller.}
\State $\varepsilon \leftarrow \min(\varepsilon,T-(t+\tau))$
\Comment{No simulation beyond process time $t=T$.}
\State $\hat{\mathbf s}(\tau+\varepsilon) = \psi_\varepsilon (\hat{\mathbf s}(\tau))$ for ODE (\ref{eq:augmented_sec_order_ode}) 
\Comment{The RKN step}
\State $\xi\leftarrow 1$
\If{$\hat{\mathbf s}_{2d+1}(\tau+\varepsilon) > u$}
\Comment{"If(event occurred during this integration step)"}
\State Find $\xi$ so that $\hat{\mathbf s}_{2d+1}(\tau+\xi \varepsilon) = u$ 
\Comment{Time between-events was $\tau+\xi \varepsilon$.}
\EndIf
\State $\mathcal Q(t+\tau+r) \leftarrow \hat{\mathbf s}_{1:d}(\tau+r),\;r\in [0,\xi \varepsilon)$ 
\Comment{Position $\mathbf q$.}
\State $\mathcal P (t+\tau+r) \leftarrow \mathbf M \hat{\mathbf s}_{d+1:2d}(\tau+r),\;r\in [0,\xi \varepsilon)$
\Comment{Momentum $\mathbf p$.}
\While{$\tau+\xi \varepsilon \geq iT/N$} 
\Comment{The $i$th sample during current integration step?}
\State $\mathbf S_{1:d,i} \leftarrow \mathcal Q(iT/N)$
\Comment{Collect sample (in practice done using RKN step interpolation).}
\State $i \leftarrow i+1$
\Comment{Advance discrete time samples counter.}
\EndWhile
\State $\tau \leftarrow \tau+\varepsilon$
\Comment{Update Hamiltonian dynamics time}
\EndWhile \Comment{Hamiltonian dynamics integration loop}
\State $t\leftarrow t+ (\tau-\varepsilon)+\xi \varepsilon$.
\Comment{Update process time $t$}
\State $\mathbf g \leftarrow \mathbf g + \hat{\mathbf s}_{2d+2:2d+1+m}((\tau-\varepsilon)+\xi \varepsilon)$
\Comment{$t^{-1}\mathbf g$ estimates moments}
\EndWhile \Comment{Event loop}
\State
\State {\bf Return} $T^{-1}\mathbf g$ 
\Comment{$T^{-1}\mathbf g_k \rightarrow \int g_k(\mathbf q)\pi(\mathbf q)d\mathbf q$ as $T\rightarrow \infty.$} 
\State {\bf Return} $\mathbf S$.
\Comment{The columns of $\mathbf S$ are dependent random draws whose marginal distribution approaches $\pi(\mathbf q)$.}
\end{algorithmic}\caption{\label{alg:Basic-CTHMC-algorithm}Basic NGRHMC algorithm based on
modified RKN integrator. Notice, this formulation assumes that the
event intensity $\lambda$ depends on $\mathbf{q}$ only. The algorithm
produces both continuous time estimates (\ref{eq:integrated-trajectory})
of the given moments, and discrete time samples $\mathbf{S}$ which
may be used similarly to conventional MCMC samples.}
\end{algorithm}
A quite general outline of the numerical algorithm for simulating
NGRHMC processes is given in Algorithm \ref{alg:Basic-CTHMC-algorithm}.
The outlined algorithm produces both integrated moments approximating
$\int g_{k}(\mathbf{q})\pi(\mathbf{q})d\mathbf{q},\;k=1,\dots,m$
based on (\ref{eq:integrated-trajectory}) and discrete time samples
$\mathbf{S}$. Notice that Algorithm \ref{alg:Basic-CTHMC-algorithm}
assumes event intensities not depending on the momentum coordinate.
However, it is straight forward to extend to the Algorithm to account
for momentum-dependent event-rates based on the interpolation option
of the RKN step.

\section{The salamander mating data using crossed random effects\label{sec:The-salamander-mating}}

\begin{table}
\centering{}%
\begin{tabular}{lccccccccccc}
\hline 
 & \begin{turn}{90}
{\footnotesize{}sampling}
\end{turn} & \begin{turn}{90}
{\footnotesize{}Prec$(\mathbf{b}_{i1}^{F})$}
\end{turn} & \begin{turn}{90}
{\footnotesize{}Prec$(\mathbf{b}_{i2}^{F})$}
\end{turn} & \begin{turn}{90}
{\footnotesize{}corr($\mathbf{b}_{i1}^{F},\mathbf{b}_{i2}^{F})$}
\end{turn} & \begin{turn}{90}
{\footnotesize{}Prec$(\mathbf{b}_{i1}^{M})$}
\end{turn} & \begin{turn}{90}
{\footnotesize{}Prec$(\mathbf{b}_{i2}^{M})$}
\end{turn} & \begin{turn}{90}
{\footnotesize{}corr($\mathbf{b}_{i1}^{M},\mathbf{b}_{i2}^{M})$}
\end{turn} & {\footnotesize{}$\tau_{F}$} & {\footnotesize{}$\tau_{M}$} & {\footnotesize{}$\mathbf{b}_{\cdot\cdot}^{\cdot}$} & {\footnotesize{}$\beta$}\tabularnewline
\hline 
\multicolumn{12}{c}{{\footnotesize{}rstan, total sampling CPU time: 23.43 s ($\max\hat{R}=1.005$)}}\tabularnewline
\hline 
{\footnotesize{}post. mean} &  & {\footnotesize{}1.09} & {\footnotesize{}0.95} & {\footnotesize{}-0.08} & {\footnotesize{}1.55} & {\footnotesize{}1.11} & {\footnotesize{}0.63} & {\footnotesize{}2.19} & {\footnotesize{}0.74} &  & \tabularnewline
{\footnotesize{}post. sd} &  & {\footnotesize{}0.94} & {\footnotesize{}0.79} & {\footnotesize{}0.40} & {\footnotesize{}1.20} & {\footnotesize{}0.89} & {\footnotesize{}0.26} & {\footnotesize{}1.51} & {\footnotesize{}0.59} &  & \tabularnewline
{\footnotesize{}ESS/CPU time} &  & {\footnotesize{}241} & {\footnotesize{}217} & {\footnotesize{}103} & {\footnotesize{}194} & {\footnotesize{}226} & {\footnotesize{}137} & {\footnotesize{}356} & {\footnotesize{}235} & {\footnotesize{}$\geq$243} & {\footnotesize{}$\geq$178}\tabularnewline
\hline 
\multicolumn{12}{c}{{\footnotesize{}pdphmc, event specification 1, $\gamma=3$, total
CPU time: 15.19 s ($\max\hat{R}=1.010$)}}\tabularnewline
\hline 
{\footnotesize{}post. mean} & {\footnotesize{}D} & {\footnotesize{}1.07} & {\footnotesize{}0.92} & {\footnotesize{}-0.09} & {\footnotesize{}1.52} & {\footnotesize{}1.09} & {\footnotesize{}0.64} & {\footnotesize{}2.13} & {\footnotesize{}0.68} &  & \tabularnewline
{\footnotesize{}post. sd} & {\footnotesize{}D} & {\footnotesize{}0.91} & {\footnotesize{}0.77} & {\footnotesize{}0.40} & {\footnotesize{}1.18} & {\footnotesize{}0.84} & {\footnotesize{}0.26} & {\footnotesize{}1.52} & {\footnotesize{}0.49} &  & \tabularnewline
{\footnotesize{}ESS/CPU time} & {\footnotesize{}D} & {\footnotesize{}149} & {\footnotesize{}188} & {\footnotesize{}155} & {\footnotesize{}274} & {\footnotesize{}244} & {\footnotesize{}200} & {\footnotesize{}333} & {\footnotesize{}252} & {\footnotesize{}$\geq$319} & {\footnotesize{}$\geq$191}\tabularnewline
{\footnotesize{}ESS/CPU time} & {\footnotesize{}C} & {\footnotesize{}137} & {\footnotesize{}145} & {\footnotesize{}155} & {\footnotesize{}246} & {\footnotesize{}156} & {\footnotesize{}201} & {\footnotesize{}352} & {\footnotesize{}159} & {\footnotesize{}$\geq$339} & {\footnotesize{}$\geq$198}\tabularnewline
\hline 
\multicolumn{12}{c}{{\footnotesize{}pdphmc, event specification 1, $\gamma=10.0$, total
CPU time: 14.91 s ($\max\hat{R}=1.015$)}}\tabularnewline
\hline 
{\footnotesize{}post. mean} & {\footnotesize{}D} & {\footnotesize{}1.08} & {\footnotesize{}0.91} & {\footnotesize{}-0.08} & {\footnotesize{}1.55} & {\footnotesize{}1.12} & {\footnotesize{}0.62} & {\footnotesize{}2.25} & {\footnotesize{}0.74} &  & \tabularnewline
{\footnotesize{}post. sd} & {\footnotesize{}D} & {\footnotesize{}0.91} & {\footnotesize{}0.75} & {\footnotesize{}0.40} & {\footnotesize{}1.21} & {\footnotesize{}0.91} & {\footnotesize{}0.26} & {\footnotesize{}1.56} & {\footnotesize{}0.55} &  & \tabularnewline
{\footnotesize{}ESS/CPU time} & {\footnotesize{}D} & {\footnotesize{}244} & {\footnotesize{}316} & \textbf{\footnotesize{}164} & {\footnotesize{}339} & {\footnotesize{}313} & {\footnotesize{}206} & {\footnotesize{}344} & {\footnotesize{}320} & {\footnotesize{}$\geq$326} & {\footnotesize{}$\geq$215}\tabularnewline
{\footnotesize{}ESS/CPU time} & {\footnotesize{}C} & {\footnotesize{}142} & {\footnotesize{}326} & \textbf{\footnotesize{}164} & \textbf{\footnotesize{}355} & \textbf{\footnotesize{}321} & {\footnotesize{}207} & \textbf{\footnotesize{}365} & \textbf{\footnotesize{}329} & {\footnotesize{}$\geq$347} & {\footnotesize{}$\geq$}\textbf{\footnotesize{}224}\tabularnewline
\hline 
\multicolumn{12}{c}{{\footnotesize{}pdphmc, event specification 2, $\gamma=3.0$, total
CPU time: 14.69 s ($\max\hat{R}=1.014$)}}\tabularnewline
\hline 
{\footnotesize{}post. mean} & {\footnotesize{}D} & {\footnotesize{}1.09} & {\footnotesize{}0.91} & {\footnotesize{}-0.08} & {\footnotesize{}1.55} & {\footnotesize{}1.13} & {\footnotesize{}0.63} & {\footnotesize{}2.14} & {\footnotesize{}0.73} &  & \tabularnewline
{\footnotesize{}post. sd} & {\footnotesize{}D} & {\footnotesize{}0.91} & {\footnotesize{}0.74} & {\footnotesize{}0.40} & {\footnotesize{}1.17} & {\footnotesize{}0.89} & {\footnotesize{}0.26} & {\footnotesize{}1.50} & {\footnotesize{}0.56} &  & \tabularnewline
{\footnotesize{}ESS/CPU time} & {\footnotesize{}D} & {\footnotesize{}265} & {\footnotesize{}279} & {\footnotesize{}160} & {\footnotesize{}246} & {\footnotesize{}316} & {\footnotesize{}209} & {\footnotesize{}339} & {\footnotesize{}189} & {\footnotesize{}$\geq$330} & {\footnotesize{}$\geq$204}\tabularnewline
{\footnotesize{}ESS/CPU time} & {\footnotesize{}C} & {\footnotesize{}208} & {\footnotesize{}186} & {\footnotesize{}161} & {\footnotesize{}221} & {\footnotesize{}250} & \textbf{\footnotesize{}210} & {\footnotesize{}357} & {\footnotesize{}128} & {\footnotesize{}$\geq$}\textbf{\footnotesize{}352} & {\footnotesize{}$\geq$209}\tabularnewline
\hline 
\multicolumn{12}{c}{{\footnotesize{}pdphmc, event specification 2, $\gamma=10.0$, total
CPU time: 15.25 s ($\max\hat{R}=1.018$)}}\tabularnewline
\hline 
{\footnotesize{}post. mean} & {\footnotesize{}D} & {\footnotesize{}1.09} & {\footnotesize{}0.93} & {\footnotesize{}-0.08} & {\footnotesize{}1.53} & {\footnotesize{}1.13} & {\footnotesize{}0.62} & {\footnotesize{}2.22} & {\footnotesize{}0.71} &  & \tabularnewline
{\footnotesize{}post. sd} & {\footnotesize{}D} & {\footnotesize{}0.90} & {\footnotesize{}0.77} & {\footnotesize{}0.40} & {\footnotesize{}1.15} & {\footnotesize{}0.88} & {\footnotesize{}0.27} & {\footnotesize{}1.59} & {\footnotesize{}0.54} &  & \tabularnewline
{\footnotesize{}ESS/CPU time} & {\footnotesize{}D} & \textbf{\footnotesize{}270} & {\footnotesize{}318} & {\footnotesize{}159} & {\footnotesize{}322} & {\footnotesize{}297} & {\footnotesize{}198} & {\footnotesize{}338} & {\footnotesize{}310} & {\footnotesize{}$\geq$321} & {\footnotesize{}$\geq$208}\tabularnewline
{\footnotesize{}ESS/CPU time} & {\footnotesize{}C} & {\footnotesize{}198} & \textbf{\footnotesize{}327} & {\footnotesize{}159} & {\footnotesize{}271} & {\footnotesize{}254} & {\footnotesize{}200} & {\footnotesize{}359} & {\footnotesize{}132} & {\footnotesize{}$\geq$342} & {\footnotesize{}$\geq$216}\tabularnewline
\hline 
\end{tabular}\caption{\label{tab:salamander}Results for the Salamander mating experiment.
The table gives posterior means (post. mean), posterior standard deviations
(post. sd.), and also the number of effective samples produced per
second of computing time (ESS/CPU time). For rstan, the results are
based on 10 independent chains with 1000 transitions proceeding 1000
warmup transitions. For pdphmc, the results are based on trajectories
of length $T=2,000$ with the first half discarded as warmup and VARI
mass. The post warmup period was discretely (D) sampled $N=1000$
times and also continuous (C) samples were recorded. The former 8
columns focus on the posterior random effects precision structure,
where (unconditional) precisions are denoted by ``Prec'' and correlations
by ``corr''. The last two columns give the worst case time-weighted
ESSes for the random- and fixed effects respectively. The best overall
sampling efficiencies are indicated with \textbf{bold} font.}
\end{table}
For the purpose of further benchmarking of the proposed methodology
against rstan, this section considers a crossed random effects model
for the Salamander mating data of \citet[chap 14.5]{mccu:neld:1989}.
The formulation of the model is an example model for INLA (see Salamander
Model B at \href{http://www.r-inla.org/examples/volume-ii}{http://www.r-inla.org/examples/volume-ii}),
the particular parameterization is taken from \citet[Section 5.3, "Full CIP/DRHMC"]{doi:10.1080/10618600.2019.1584901}
and the rstan implementation is identical to that of \citet{doi:10.1080/10618600.2019.1584901}.
The model is characterized by random effects precision priors
\begin{align*}
\tau_{F} & \sim\text{Gamma}(1,0.622),\;\tau_{M}\sim\text{Gamma}(1,0.622),\\
\mathbf{P}_{F} & \sim\text{Wishart}_{2}(3.0,\mathbf{W}),\;\mathbf{P}_{M}\sim\text{Wishart}_{2}(3.0,\mathbf{W}),\;\mathbf{W}=\text{diag}(0.804,0.804),
\end{align*}
random effects
\begin{align}
[\mathbf{b}_{i1}^{F},\mathbf{b}_{i2}^{F}]^{T} & \sim\text{iid }N(\mathbf{0}_{2},\mathbf{P}_{F}^{-1}),\;i=1,\dots,20,\label{eq:salamander_re1}\\{}
[\mathbf{b}_{j1}^{M},\mathbf{b}_{j2}^{M}]^{T} & \sim\text{iid }N(\mathbf{0}_{2},\mathbf{P}_{M}^{-1}),\;j=1,\dots,20,\label{eq:salamander_re2}\\
\mathbf{b}_{i3}^{F} & \sim\text{iid }N(0,\tau_{F}^{-1}),\;i=1,\dots,20,\nonumber \\
\mathbf{b}_{j3}^{M} & \sim\text{iid }N(0,\tau_{M}^{-1}),\;j=1,\dots,20,\nonumber 
\end{align}
and finally the observation equation
\[
\mathbf{y}_{ijk}|\boldsymbol{\Pi}_{ijk}\sim\text{Bernoulli}(\boldsymbol{\Pi}_{ijk}),\;\text{logit}(\boldsymbol{\Pi}_{ijk})=\mathbf{x}_{ijk}^{T}\beta+\mathbf{b}_{ik}^{F}+\mathbf{b}_{jk}^{F}.
\]
The observations are coded as 1 for successful mating of female salamander
$i$ and male salamander $j$ in experiment $k=1,2,3$. The salamanders
in experiment $k=1,2$ are identical, and therefore their associated
random effects (\ref{eq:salamander_re1},\ref{eq:salamander_re2})
are allowed to be correlated. Finally, $\mathbf{x}_{ijk}\in\mathbb{R}^{5}$
is a covariate vector and $\beta\in\mathbb{R}^{5}$ is a fixed effect
with flat prior. Experiment outcomes were recorded in 360 combinations
of $i,j,k$. 

Details on the parameterization of model, aiming at making the target
distribution suitable for HMC sampling by reducing ``funnel'' effects,
are given in \citet{doi:10.1080/10618600.2019.1584901}. Still the
target distribution is quite far from being Gaussian. In total, the
model involves $d=133$ free parameters and random effects to be sampled. 

Table \ref{tab:salamander} provides results, focusing on the posterior
marginal precisions and correlations of random effects, as these are
typically rather difficult to estimate using MCMC. The overall picture
from the experiment is that pdphmc is slightly more efficient than
rstan.

\section{Dynamic inverted Wishart model details\label{subsec:details__dynamic-inverted-Wishart}}

\begin{table}
\centering{}%
\begin{tabular}{ccccccccc}
\hline 
 & \multicolumn{2}{c}{rstan} &  & \multicolumn{2}{c}{pdphmc} &  & \multicolumn{2}{c}{pdphmc}\tabularnewline
 &  &  &  & \multicolumn{2}{c}{event spec. 1} &  & \multicolumn{2}{c}{event spec. 3}\tabularnewline
\cline{2-3} \cline{3-3} \cline{5-6} \cline{6-6} \cline{8-9} \cline{9-9} 
 & post.  & post.  &  & post.  & post.  &  & post.  & post. \tabularnewline
 & mean & SD &  & mean & SD &  & mean & SD\tabularnewline
\hline 
$\boldsymbol{\mu}_{1}$  & 4.15  & 0.199  &  & 4.16  & 0.206  &  & 4.16  & 0.202 \tabularnewline
$\boldsymbol{\mu}_{2}$  & 4.12  & 0.263  &  & 4.12  & 0.244  &  & 4.12  & 0.268 \tabularnewline
$\boldsymbol{\mu}_{3}$  & 3.71  & 0.154  &  & 3.72  & 0.149  &  & 3.72  & 0.154 \tabularnewline
$\boldsymbol{\mu}_{4}$  & 4.11  & 0.095  &  & 4.11  & 0.092  &  & 4.11  & 0.092 \tabularnewline
$\boldsymbol{\mu}_{5}$  & 3.53  & 0.136  &  & 3.53  & 0.134  &  & 3.53  & 0.132 \tabularnewline
$\boldsymbol{\sigma}_{1}$  & 0.31  & 0.009  &  & 0.31  & 0.009  &  & 0.31  & 0.009 \tabularnewline
$\boldsymbol{\sigma}_{2}$  & 0.26  & 0.008  &  & 0.26  & 0.008  &  & 0.26  & 0.008 \tabularnewline
$\boldsymbol{\sigma}_{3}$  & 0.29  & 0.009  &  & 0.29  & 0.009  &  & 0.29  & 0.009 \tabularnewline
$\boldsymbol{\sigma}_{4}$  & 0.28  & 0.009  &  & 0.28  & 0.009  &  & 0.28  & 0.009 \tabularnewline
$\boldsymbol{\sigma}_{5}$  & 0.25  & 0.009  &  & 0.25  & 0.009  &  & 0.25  & 0.009 \tabularnewline
$\boldsymbol{\delta}_{1}$  & 0.97  & 0.005  &  & 0.97  & 0.005  &  & 0.97  & 0.005 \tabularnewline
$\boldsymbol{\delta}_{2}$  & 0.98  & 0.004  &  & 0.98  & 0.004  &  & 0.98  & 0.004 \tabularnewline
$\boldsymbol{\delta}_{3}$  & 0.96  & 0.006  &  & 0.96  & 0.006  &  & 0.96  & 0.006 \tabularnewline
$\boldsymbol{\delta}_{4}$  & 0.94  & 0.008  &  & 0.94  & 0.008  &  & 0.94  & 0.008 \tabularnewline
$\boldsymbol{\delta}_{5}$  & 0.96  & 0.006  &  & 0.96  & 0.006  &  & 0.96  & 0.006 \tabularnewline
$\mathbf{H}_{2,1}$  & 0.39  & 0.003  &  & 0.39  & 0.003  &  & 0.39  & 0.003 \tabularnewline
$\mathbf{H}_{3,1}$ & 0.29  & 0.003  &  & 0.29  & 0.003  &  & 0.29  & 0.003 \tabularnewline
$\mathbf{H}_{4,1}$  & 0.29  & 0.003  &  & 0.29  & 0.003  &  & 0.29  & 0.003 \tabularnewline
$\mathbf{H}_{5,1}$  & 0.23  & 0.002  &  & 0.23  & 0.002  &  & 0.23  & 0.002 \tabularnewline
$\mathbf{H}_{3,2}$  & 0.20  & 0.003  &  & 0.20  & 0.003  &  & 0.20  & 0.003 \tabularnewline
$\mathbf{H}_{4,2}$  & 0.17  & 0.003  &  & 0.17  & 0.003  &  & 0.17  & 0.003 \tabularnewline
$\mathbf{H}_{5,2}$  & 0.12  & 0.002  &  & 0.12  & 0.002  &  & 0.12  & 0.002 \tabularnewline
$\mathbf{H}_{4,3}$  & 0.22  & 0.004  &  & 0.22  & 0.004  &  & 0.22  & 0.004 \tabularnewline
$\mathbf{H}_{5,3}$  & 0.18  & 0.003  &  & 0.18  & 0.003  &  & 0.18  & 0.003 \tabularnewline
$\mathbf{H}_{5,4}$  & 0.11  & 0.002  &  & 0.11  & 0.002  &  & 0.11  & 0.002 \tabularnewline
$\nu$  & 33.61  & 0.291  &  & 33.61  & 0.289  &  & 33.61  & 0.278 \tabularnewline
\hline 
\end{tabular}\caption{\label{tab:Posterior-means-and-wishart}Posterior means and standard
deviations for the parameters of the dynamic inverted Wishart model
of Section \ref{subsec:Dynamic-Inverted-Wishart}. Only results based
on discrete samples are presented for the pdphmc methods, as the continuous
samples means are identical to the discrete samples means in the precision
reported here.}
\end{table}
The priors and transformations to a unrestricted domain for the parameters
are given by
\begin{itemize}
\item $\boldsymbol{\delta}_{g}\sim\text{iid }Uniform(-1,1),\;g=1,\dots,G$.
Sampling was performed in a scaled logit-transformed parameter.
\item $\boldsymbol{\mu}_{g}\sim\text{\text{iid }}N(0,5^{2}),\;g=1,\dots,G.$
No transformation applied.
\item $\boldsymbol{\sigma}_{g}^{2}\sim\text{iid }p_{0}s_{0}/\chi_{p_{0}}^{2},\;g=1,\dots,G$,
where $p_{0}=4$ and $s_{0}=0.25$. Sampling was performed in $\log(\sigma_{g})$.
\item $\mathbf{H}_{i,j}\sim\text{iid }N(0,10^{2}),\;j=1,\dots,G-1,i=j+1,\dots,G$.
No transformation applied.
\item $\nu\sim Uniform(10.0,60.0)$, Sampling was performed in a scaled
logit-transformed parameter.
\end{itemize}
A fortunate property of this model, owing to the particular choice
of observation equation (\ref{eq:wishart_1}) and scale matrix (\ref{eq:wishart_2})
is that $p(\mathbf{x}|\mathbf{Y}_{1:n},\boldsymbol{\theta})$ factorizes
as $\prod_{g=1}^{G}p(\mathbf{x}_{g,1:n}|\boldsymbol{\theta})h(\mathbf{x}_{g,1:n})$,
i.e. the latent factors are conditionally on $\mathbf{Y}_{1:n},\boldsymbol{\theta}$
independent over $g$. Moreover the function $h(\mathbf{x}_{g,1:n})$
has a particularly simple form
\[
h(\mathbf{x}_{g,1:n})\propto\prod_{k=1}^{n}\exp\left[\frac{\nu}{2}\mathbf{x}_{g,k}-\frac{\tilde{\boldsymbol{y}}_{g,k}}{2}\exp(\mathbf{x}_{g,k})\right],\;\tilde{\boldsymbol{y}}_{g,k}=\left[\mathbf{H}_{1:G,g}\right]^{T}\mathbf{Y}_{k}^{-1}\left[\mathbf{H}_{1:G,g}\right].
\]
I.e. $p(\mathbf{x}|\mathbf{Y}_{1:n},\boldsymbol{\theta})$ factorizes
into the form of $G$ independent (conditional on $\boldsymbol{\theta}$)
smoothing distributions associated with single-dimensional state space
models.

The transport map \citep[see][for more details]{1812.07929} $\gamma_{\boldsymbol{\theta}}$
relating $\mathbf{x}$ and $\mathbf{z}$ is given by 
\[
\mathbf{x}_{g,1:n}=\mathbf{h}_{g,\boldsymbol{\theta}}+\mathbf{L}_{g,\boldsymbol{\theta}}^{-T}\mathbf{z}_{(g-1)n+1:gn},\;g=1,\dots,G
\]
where $\mathbf{L}_{g,\boldsymbol{\theta}}$ is the lower Cholesky
triangle of the SPD tri-diagonal matrix
\[
\mathbf{G}_{g,\boldsymbol{\theta}}=\text{\ensuremath{\boldsymbol{G}}}_{g,\mathbf{x}|\boldsymbol{\theta}}+\mathbf{G}_{g,\mathbf{y}|\mathbf{x},\boldsymbol{\theta}}.
\]
Here $\text{\ensuremath{\boldsymbol{G}}}_{g,\mathbf{x}|\boldsymbol{\theta}}$
is the (tri-diagonal) precision matrix of $\mathbf{x}_{g,1:n}|\boldsymbol{\theta}$
and $\mathbf{G}_{g,\mathbf{y}|\mathbf{x},\boldsymbol{\theta}}$ is
the (diagonal) negative Hessian of $\mathbf{x}_{g,1:n}\mapsto\log(h(\mathbf{x}_{g,1:n}))$
evaluated at $\mathbf{h}_{g,\mathbf{y}|\mathbf{x},\boldsymbol{\theta}}=\arg\max_{\mathbf{x}_{g,1:n}}\log(h(\mathbf{x}_{g,1:n}))$
(this optimization problem splits into $n$ separate single variable
problems and is analytically solved). Finally 
\[
\mathbf{h}_{g,\boldsymbol{\theta}}=\mathbf{G}_{g,\boldsymbol{\theta}}^{-1}\left[\text{\ensuremath{\boldsymbol{G}}}_{g,\mathbf{x}|\boldsymbol{\theta}}E(\mathbf{x}_{g,1:n}|\boldsymbol{\theta})+\mathbf{G}_{g,\mathbf{y}|\mathbf{x},\boldsymbol{\theta}}\mathbf{h}_{g,\mathbf{y}|\mathbf{x},\boldsymbol{\theta}}\right].
\]
The Jacobian determinant of $\text{vec}(\gamma_{\boldsymbol{\theta}}(\mathbf{z}))$
reduces to $\prod_{g=1}^{G}|\mathbf{L}_{g,\boldsymbol{\theta}}^{-T}|$.

\section{Further work\label{sec:Further-work}}

A, by no means complete, list of possible further research directions
related to NGRHMC processes contains the following:
\begin{itemize}
\item More adaptive event specifications in a similar vein to those of \citet{JMLR:v15:hoffman14a,1601.00225}
to obtain trajectory lengths that are better adapted to local geometry
of the target distribution. In this regard, working with multiple
independent copies of target distribution, with negatively correlated
momentums (across the copies) may be useful. 
\item The results of Section \ref{sec:Continuous-time-HMC} goes through
with minimal changes for non-Euclidian Riemann manifold Hamiltonian
dynamics \citep{girolami_calderhead_11} (see Appendix \ref{sec:Riemann-manifold-variants}).
In this case, the dynamics solve a coupled system of $2d$ first order
ODEs (but cannot be reduced to a $d$-dimensional second order system).
In particular, such a development could be carried out with explicit
(non-symplectic) numerical integrators for first order systems, which
could substantially reduce the often large computational cost of conventional
Riemann manifold HMC \citep[see e.g.][]{Kleppe2017}. 
\item Recently, pseudo-marginal/transport map MCMC methods \citep{Lindsten2016,doi:10.1080/10618600.2019.1584901,1812.07929}
for large scale Bayesian hierarchical models have been proposed. Such
methods result in a modified target distribution which ``almost''
factorizes into a low-dimensional general factor and a high-dimensional
standard Gaussian factor. A symplectic integrator optimized for such
a situation was developed in \citet{Lindsten2016}. A further development
in this project would be to develop a non-symplectic integrator optimized
for such ``almost factorizing'' situations, e.g., using splitting
methods \citep[see e.g.][]{FRANCO2014482}.
\item It may be beneficial to operate with multiple event rates/momentum
refreshes for different parts of the state vector. Theoretically,
this should be rather straight forward, but finding good strategies
for such a development would require substantial work. 
\item Reversible variants (based on involutions and reversible jump Metropolis
Hastings) using the RKN methodology leads to approximate discrete
time MCMC algorithms inheriting many of the properties of NGRHMC related
to computational robustness. Such algorithms are attractive in that
they allow very flexible adaptive selection of the Hamiltonian trajectory
(time-)length distribution (e.g., similarly to NUTS). However, their
efficient computation requires storing a representation of complete
Hamiltonian trajectories and may involve simulation of substantial
amounts of dynamics which is only used for calculating accept probabilities
(both as opposed to the PDMP version).
\item A further interesting property of NGRHMC processes is that biased
(relative to the chosen momentum target distribution $N(\mathbf{0}_{d},\mathbf{M})$)
updates can be corrected by appropriately chosen event rates. This
may be exploited e.g., toward higher momentums intended for jumps
between modes, and implemented in a stable manner using the adaptive
integrators.
\end{itemize}

\section{Simple R-implementation\label{sec:Simple-R-implementation}}

The computations detailed in the paper are based on a rather complicated
code written in C++ for best performance. The purpose of this section
is to show that working implementations of NGRHMC processes for smaller
problems or prototypes can also be implemented very easily in high
level languages with access to off-the-shelf (first order) ODE solvers.
Below is an R-implementation of the proposed methodology based on
the ODE code \texttt{lsodar} (from package \texttt{deSolve}) which
supports root-finding/event capabilities. Such capabilities may be
leveraged to run the whole simulation as single call to the solver. 

The below code considers a zero mean bivariate target distribution
with unit marginal variances and correlation 0.5. Other (bivariate)
target distributions are obtained by simply changing the function
\texttt{target.grad} accordingly. A constant $\lambda=0.1$ and identity
mass matrix $\mathbf{M}$ was used. Since \texttt{lsodar} is a first
order ODE code, the representation (\ref{eq:Ham_eq}) of Hamilton's
equations is used, and in addition $\Lambda$ and continuous time
temporal averages are added to the system of ODEs.

\begin{lstlisting}[language=R]
# Simple implementation of CTHMC based on the
# (first order ODE) solver lsodar provided by
# package "deSolve". The mass matrix is implicitly
# taken to be the identity, and a constant event rate
# is considered
set.seed(1)
# constants
lambda <- 0.1 # rather infrequent events
T <- 10000.0 # total simulation time
N.samples <- 2000 #How many uniformly spaced samples of the process

# consider a N(0,Sigma) target where
Sigma <- matrix(c(1,0.5,0.5,1),nrow=2)
prec <- solve(Sigma) # precision matrix
#target gradient (change to get another target distribution)
target.grad <- function(q){ return(-prec%*%q)}

# define the (first order ODE) for state
# state[1:2] : q
# state[3:4] : \dot{q} = p
# state[5] : \Lambda
# state[6] : u
# state[7:8] : \int q dt

ode <- function(t,state,parms){
  ret <- c(state[3:4], #\dot q = p
           target.grad(state[1:2]), # \dot p = gradient of log-target
           lambda, # \dot \Lambda = \lambda
           0.0, # u is constant
           state[1:2]) # integrated quantities
  return(list(ret)) #ODE solver assumes a list
}
#
# events occur when this function evaluate to 0: here \Lambda=u
#
root.fun <- function(t,state,parms){return(state[5]-state[6])}
#
# what happens at events?
#
at.event <- function(t,state,parms){
  newState <- c(state[1:2], # q unchanged
                rnorm(2), # p refreshed
                0.0, # reset Lambda
                rexp(1), # simulate new u
                state[7:8]) # continue integrating
  return(newState)
}

# initial state, starting from q=(1,1)
y0 <- c(1,1,# q(0)
        rnorm(2), #p(0)
        0.0, #Lambda
        rexp(1), #u
        0.0,0.0) #integrated quantities
#
# run actual simulation
#
sim.out <- deSolve::lsodar(y=y0,
                           func=ode,
                           times=seq(from=0,to=T,length.out = N.samples+1),
                           rootfunc = root.fun,
                           events = list(func=at.event,root=TRUE))

# display q_1 discrete samples
par(mfrow=c(1,2))
ts.plot(sim.out[,"1"])
acf(sim.out[,"1"])

# print summary stats
print(summary(sim.out[,c("1","2")]))
print(cor(sim.out[,c("1","2")]))

# construct integrated samples
eta.1 <- diff(sim.out[,"7"])*(N.samples/T)
# display integrated samples
ts.plot(eta.1)
acf(eta.1)
mean(eta.1) # estimator of E(q_1)
sd(eta.1) # < SD(q_1)




\end{lstlisting}
\end{document}